\documentstyle[preprint,aps,epsf]{revtex}
\begin{document}
\draft

\title{EFFECTIVE SURFACE IMPEDANCE OF POLYCRYSTALS
UNDER CONDITIONS OF ANOMALOUS SKIN-EFFECT}

\author{Inna M. Kaganova}
\address{Institute of High Pressure Physics Russian Academy of Sciences
Russia, 142092 Troitsk Moscow region, e-mail: u10740@dialup.podolsk.ru}

\author{Moisey I. Kaganov}
\address{7 Agassiz Ave., Belmont, MA 02478, USA, e-mail:
MKaganov@compuserve.com} 

\date{\today}
\maketitle

\begin{abstract}
The effective impedance of strongly anisotropic polycrystals has been
investigated under the conditions of extremely anomalous skin effect. We were
interested in finding out how the value of the effective impedance depends on
the geometry of the Fermi surface of a single crystal grain. The previously
obtained nonperturbative solution based on the application of the impedance
(the Leontovich) boundary conditons was used to calculate the effective
impedance of a polycrystalline metal. When the Fermi surface is a surface of
revolution, the expression for the effective impedance is rather simple. Some
model Fermi surfaces were examined. In the vicinity of the electronic
topological transition the singularities of the effective impedance related to
the change of the topology of the Fermi surface were calculated. Our results
show that though a polycrystal is an isotropic medium in average, it is not
sufficient to consider it as a metal with an effective spherical Fermi surface,
since this can lead to the loss of some characteristic features of extremely
anomalous skin effect in polycrystals.
\end{abstract}


\section{INTRODUCTION}
In recent years there has been considerable interest on the part of theorists
and experimentalists in the study of macroscopic properties of inhomogeneous 
solids. A special, but very widespread case of an inhomogeneous solid medium 
is a polycrystal whose inhomogeneity is due to the misorientation of discrete
single crystal grains. Since polycrystals are the usual state of crystalline
media, the calculation of their properties is an actual problem.

Macroscopic properties of polycrystalline solids can be described in the 
framework of different models of an effective isotropic medium. The problem is
to calculate characteristics of such an isotropic medium when the corresponding
parameters of single crystalline grains are known. For example, let 
${\sigma}_{i}$ $ (i=1,2,3)$ be the principal values of the single crystal
static conductivity tensor. By proceeding from this information we would like
to calculate the scalar effective conductivity ${\sigma}_{ef}$ of the
polycrystal. 

In our opinion the most accurate and physically meaningful method of
calculation of an effective characteristic of a polycrystal goes back to the
pioneering works of I.M.Lifshitz et al \cite{1,2,3}. In the framework of this
method it is assumed that the polycrystalline medium can be described as an
effective isotropic medium that is perturbed by random spatial fluctuations
caused by the orientational fluctuations of the grains. Generally, an effective
characteristic of a polycrystal is not a function of its value in the single
crystal only (in our example ${\sigma}_{ef}$ is not a function of
${\sigma}_{i}$ only), but depends on the geometric statistics of the grains,
reflecting directly the macroscopic properties of the medium they compose.
The statistical properties of the medium are described by the multipoint
correlators of the characteristic. These correlators, in turn, depend on the
shape, the mutual location, and the misorientation of crystallographic axes of
constituent single crystal grains. 

For the sake of being definite, let's assume that $\hat\psi$ is any tensor
macroscopic characteristic of a polycrystal. Let $\bf f$ be a vector field.
We have equations, which are associated with this field, and the characteristic
$\hat\psi$ is a coefficient of these equations (for example, if $\hat\psi$ is
the conductivity tensor, $\bf f$ is the electromagnetic field and the
differential equations are Maxwell's equations). Let us be interested in
the calculation of the field $<{\bf f}>$, which is the field $\bf f$ averaged
over realizations of the polycrystalline structure. Let's measure the elements
of the tensor $\hat\psi$ in the laboratory coordinate system. In each single
crystal grain their values depend on the orientation of the crystallographic
axes of this particular grain with respect to the laboratory axes. Thus, in the
polycrystal the elements of the tensor $\hat\psi$ are functions of position
$\bf x$ in the medium. If the crystallographic axes of the grains are rotated
randomly with respect to each other, $\hat\psi ({\bf x})$ is a stochastic
tensor. Our problem is to define the effective characteristic of the medium
${\hat\psi}_{ef}$, which provides the correct value of the averaged field
$<{\bf f}>$. 

In the general case, when calculating ${\hat\psi}_{ef}$, the geometric
statistics of the grains (the spatial fluctuations of $\hat\psi ({\bf x})$) can
be taken into account accurately only under the assumption of small anisotropy
of the characteristic $\hat\psi$ in the original single crystal. Let
$\overline{\hat\psi}$ be the tensor $\hat\psi$, whose elements are averaged
over all possible orientations of crystallites. If the spatial fluctuations
$\hat\psi ({\bf x}) - \overline{\hat\psi}$ are small, the perturbation theory
is applicable. The zero order term of the perturbation series gives 
${\hat\psi}_{ef} \approx \overline{\hat\psi}$. 

The general way providing the correct result for ${\hat\psi}_{ef}$ with regard
to its spatial fluctuations was given in \cite{1,2,3}. According to these
papers, the starting point of the calculation is the derivation of the
equations for the averaged field $<\bf f>$. To this end we present the field
$\bf f$ as a sum of the averaged field and a stochastic addition 
$\delta {\bf f}$. Averaging the exact stochastic equations and calculating the
stochastic field $\delta {\bf f}$ considering the averaged field $<\bf f>$ as
known, we obtain a system of closed equations for the averaged field. By
proceeding from these equations the value ${\hat\psi}_{ef}$ can be defined and
calculated. The correction to $\overline{\hat\psi}$ can be obtained as an
expansion in correlators $<(\hat\psi ({\bf x}_{1}) - \overline{\hat\psi})
(\hat\psi ({\bf x}_{2}) - \overline{\hat\psi})>$, $<(\hat\psi ({\bf x}_{1}) - 
\overline{\hat\psi})(\hat\psi ({\bf x}_{2}) -\overline{\hat\psi})
(\hat\psi ({\bf x}_{3}) - \overline{\hat\psi})>$ and so on; angular brackets
$<...>$ denote an average over the ensemble of realizations of the
polycrystalline structure (see, for example, ref. \cite{25}). These correlators
can be calculated for a model polycrystal, or they are assumed to be known
characteristics of the medium. Only rarely the series of the correlators can be
summed. That is the reason why, in the general case, the problem of calculation
of effective polycrystal characteristics has not been solved yet. 

We would like to note that the procedure described above, involves the
calculation of random fields $\delta \bf f$. Consequently, the calculation of
effective characteristics of an unbounded polycrystal is simpler than the 
calculation of effective characteristics related to phenomena, where the sample
surface must be taken into account. In the former case the problem is reduced
to an algebraic problem, while in the latter one an integral equation must be
solved (for details see \cite{4,5,6,7,8}). The calculation of the effective
surface impedance of a weakly anisotropic metal polycrystals carried out in the
framework of perturbation theory was presented in \cite{6,7,8}.  

The exact solutions, which allow to consider polycrystals with strong
anisotropy can be found very rarely. One of such examples is the calculation of
the effective static conductivity of a two-dimensional isotropic polycrystal,
where, due to a specific symmetry transformation allowed by the equations of
the problem, the exact result has been obtained for arbitrary values of two
principal conductivities \cite{9}. It must be pointed out that this result
does not depend on the geometric statistics of the grains (on the correlators
of the conductivity in different points of the polycrystal). 

The other example is quite a new result for the effective surface impedance of
a polycrystalline metal associated with the reflection of an averaged
electromagnetic wave \cite{10,11}. It is valid in the frequency region of the
impedance (the Leontovich) boundary conditions applicability \cite{12,13}, i.e.
when the penetration depth of electromagnetic field into a metal $\delta$ is
much less than a characteristic length $a$ related to the surface
inhomogeneity. In the case of a polycrystalline metal with the flat surface,
$a$ is of the order of the mean size of a grain.  

This result is a nonperturbative one with respect to the inhomogeneity
amplitude. It is exact up to the limits of the impedance boundary conditions 
applicability, i.e. up to the terms of the order of $\delta /a$. Although under
the conditions of strong skin effect ($\delta \ll a$) the electromagnetic
problem seems to resemble a two-dimensional problem, the structure and the
reason of the exact result \cite{10,11} existence are different from the ones
in the two-dimensional static problem. E.L.Feinberg obtained a similar result 
while calculating the effective dielectric constant for radio waves propagating
along the earth surface \cite{14}. We would like to mention that the same
approach to the effective surface impedance calculation can be used when the
surface inhomogeneity is due to the surface roughness \cite{15}.

The method of the effective impedance calculation, proposed in \cite{10,11}, is
suitable both for the conditions of normal and anomalous skin effect. The
conditions of normal skin-effect correspond to the low frequency region when
$l \ll \delta, \omega \tau \ll 1$.  Here $l$ is the electron mean free path,
$\omega$ is the frequency of the incident wave, $\tau$ is the electron
relaxation time. In this limiting case the relation between the current density
$\bf j$ and the  electric field strength $\bf E$ is local, and the conductivity
tensor is an ordinary tensor (a multiplying operator). For an isotropic metal
with a scalar conductivity $\sigma$ under the conditions of normal skin effect 
$\delta = c/\sqrt{2\pi\sigma\omega}$. 

In sufficiently clean metals, skin effect clearly shows an anomalous character
when the temperature is low and the electron mean free path $l$ exceeds the
penetration depth $\delta$. Moreover, over a wide range of frequencies of radio
waves the condition $l \gg \delta$ is easily fulfilled. This is the range of 
extremely anomalous skin effect. At the same time, the frequency $\omega$ can
be much less than $1/\tau$. Everywhere in what follows we assume that both the
inequalities ($l \gg \delta$ and $\omega\tau \ll 1$) are fulfilled. We would
like to emphasize that these conditions are not burdensome: when the
temperature is low, as a rule, anomalous skin effect takes place for radio
waves whose wavelengths range in value from centimeters to many meters. 
It is easy to verify that the inequalities $l \gg \delta$ and 
$\omega\tau \ll 1$ are consistent if
$$
{\left(\frac{{\delta}_{p}}{l}\right)}^{2} \ll \omega\tau \ll 1;\; 
{\delta}_{p} =\frac{c}{{\omega}_{p}},
$$
where ${\omega}_{p}$ is the electronic plasma frequency, given by 
${\omega}_{p}^{2} = 4\pi ne^2/{m}^{*}$; $n$ is the electron number density, $e$
is the magnitude of the electronic charge, and ${m}^{*}$ is the effective mass
of the charge carriers. Consequently, the electron mean free path $l$ must
exceed the plasma penetration depth ${\delta}_{p} \sim {10}^{-5}-{10}^{-6}{\rm
cm}$. This inequality must be taken into account when the relation between $l$
and the mean size of a grain $a$ is discussed (see below).

Under the conditions of anomalous skin effect the relation between the current
$\bf j$ and the electric field strength $\bf E$ is non-local. In this case,
Maxwell's equations turn into the system of integro-differential equations.
Let's note that in the papers of I.M.Lifshitz et al \cite{1,2,3}, as well as in
the following studies\footnote{Recently the effective characteristics of
inhomogeneous media were calculated by a lot of authors. Not pretending to
give the full list of references, for an example we cite the papers
\cite{26,27,28}.}, the calculation of effective characteristics of polycrystals
was based on the solution of differential equations with stochastic
coefficients. The study of anomalous skin effect in polycrystals appears to be
the first example of an analysis of stochastic integro-differential equations
with regard to the theory of polycrystals. 

One of distinctive features of anomalous skin effect is the possibility to pass
to the limit $l  \to \infty$. In particular this means that under the
conditions of anomalous skin effect, the impedance does not depend on
temperature \cite{16}. 

The impedance is a macroscopic characteristic of the metal surface sensitive to
the orientation of the crystal surface with respect to its crystallographic
axes. In other words, it depends on the vector $\bf n$, which is the unit
vector directed along the normal to the metal surface. Next, under the
conditions of normal skin effect the value of the impedance of a crystal is
defined by the conductivity tensor; under the conditions of anomalous skin
effect it depends on the geometry of the Fermi surface of the metal. The
difference becomes especially evident in the case of cubic crystals. When
calculating the conductivity, a cubic crystal is the same as an isotropic
solid: the conductivity tensor degenerates into a scalar. Then under the
conditions of normal skin effect the impedance of a cubic crystal does not
depend on $\bf n$. Moreover, the impedance does not depend on the structure of
the sample: it is the same for crystal and polycrystal samples. But the Fermi
surfaces of cubic crystals are rather complex. Consequently, under the
conditions of anomalus skin effect the impedance of a cubic crystal is a
complex function of $\bf n$ defined by the structure of the Fermi surface.  

In \cite{10,11} the effective impedance of various strongly anisotropic
polycrystalline media has been calculated under the conditions of normal skin
effect. In the same works, under the conditions of anomalous skin, effect only 
polycrystals composed of single crystal grains whose Fermi surfaces were 
uniaxial ellipsoids were studied. In the present publication we have to find
out whether the effective impedance of polycrystals under the conditions of
anomalous skin effect depends on the geometry of the Fermi surface.

It is well known that the Fermi surfaces of real metals are extremely 
complex and differ significantly for different metals \cite{17,29}. Let 
$\varepsilon ({\bf p}) = {\varepsilon}_{F}$ be the equation of the Fermi
surface: ${\varepsilon}_{F}$ is the Fermi energy, $\bf p$ is the Fermi electron
momentum (or rather, its quasi-momentum). The Fermi surface is periodic in $\bf
p$. In certain metals it breaks up into individual identical surfaces, each of
which is situated in its respective cell of the reciprocal lattice. Such Fermi
surfaces are called {\it closed} Fermi surfaces. The ellipsoidal Fermi surface,
discussed in \cite{10,11}, is an example of a closed Fermi surface. In other
metals the Fermi surfaces are {\it open}, passing through the whole momentum
space. 

Due to the degeneracy (the temperature $T \ll {\varepsilon}_{F}$) only
electrons whose energy is equal to ${\varepsilon}_{F}$ take part in the metal
conductivity. They are the electrons on the Fermi surface. Under the conditions
of extremely anomalous skin effect the inequality $l \gg \delta$, or $kl \gg 1$
($\bf k$ is the electromagnetic field wave vector), selects electrons from "the
belt" ${\bf k}{\bf v}_{F} = 0$; ${\bf v}_{F} = 
\partial \varepsilon ({\bf p})/\partial {\bf p}$ is the velocity of the
electron on the Fermi surface. Other electrons are ineffective \cite{17}.  

On the other hand, when calculating the impedance of a polycrystal, an
averaging over the direction of $\bf n$ has to be done (see below). When the
direction of the normal to the metal surface changes, "the belt" moves along
the Fermi surface. Thus, the impedance of the polycrystal is defined by all the
electrons from the Fermi surface even under the conditions of extremely
anomalous skin effect. The averaging leads to an isotropization. It is usual to
think of an isotropic metal as of a metal with the shperical Fermi surface. The
question is, if this evidently model assumption is correct for the description
of anomalous skin effect in polycrystals. Our results show that being a
characteristic of a medium which is isotropic on the average, the effective
impedance of a polycrystal composed of single crystal grains with complex Fermi
surface depends on the details of the geometry of the Fermi surface.   

As it is shown below, the calculation of the effective impedance of
polycrystals involves two steps. The first step is the calculation of the
impedance of the crystal metal for an arbitrary orientation of the
crystallographic axes with respect to the metal surface. And the second step is
the averaging over all possible orientations of the crystallographic axes. The
first step requires the definition of electrons providing the maximal
contribution to the conductivity when $kl \gg 1$. It is well known, the more
the conductivity, the less the impedance. The second step (the averaging)
selects the calculated impedances choosing the maximal. Thus, the required
impedance is the result of the solution of a nontrivial mini-max problem. 

The outline of this paper is as follows. Following ref. \cite{15} in Section II
we present the algorithm of the effective surface impedance calculation
suitable in the case of strongly anisotropic polycrystals. In Section III the 
applicability of this approach to polycrystals under the conditions of
anomalous skin effect is discussed. Based on the results of Section II, we
obtain the general expression for the effective impedance of a polycrystal
metal composed of single crystal grains with a complex dispersion relation of
conduction electrons, i.e. with a complex Fermi surface. In the case of Fermi
surfaces which are arbitrary surfaces of revolution, the result is presented
in the form which allows analytical analyses of the influence of the shape of
the Fermi surface  on the effective impedance value. In Section IV we use the
obtained formulae to calculate the effective impedance for polycrystals
composed of the grains with different model Fermi surfaces. In Section V we
investigate the effect of the change of the topology of the Fermi surface on
the value of the effective impedance of polycrystals and present two examples
of its behavior in the vicinity of the electronic topological transition.  

\section{THE EFFECTIVE IMPEDANCE CALCULATION}

It is well known that to solve an electrodynamic problem external with 
respect to the metal, it is sufficient to know the tangential components of the
electric $\bf E$ and magnetic $\bf H$ vectors at the metal surface \cite{13}.
Only the relation between ${\bf E}_{t} $ and ${\bf H}_{t}$ depends on the 
metal properties:
$$
{\bf E}_{t} = \hat \zeta [{\bf n},{\bf H}_{t}],    \eqno (1)
$$
the subscript "t" denotes the tangential components of the vectors. The
two-dimensional tensor $\hat \zeta$ is the surface impedance tensor of the
metal. Its real and imaginary parts define respectively the absorption of an
incident wave and the phase shift of the reflected wave.

Because of a very high conductivity the character of the reflection of an
electromagnetic wave from a metal surface practically does not depend on the
shape of the incident field, in particular, on the angle of incidence. Then, to
calculate the surface impedance, it is sufficient to investigate normal
incidence of an electromagnetic wave onto the metal half-space. The explicit
form of the elements of $\hat\zeta$ depends on the frequency of the incident
wave $\omega$ as well as on the metal characteristics \cite{17}. In the order
of magnitude $|{\zeta}_{\alpha\beta}| \sim \delta /\lambda \ll 1$; $\lambda$ is
the vacuum wave length, $\alpha,\beta =1,2$.  
  
When the tensor $\hat\zeta$ is known, Eqs.(1) play the role of boundary 
conditions for the electromagnetic fields in the region outside the metal. If a
metal surface is an inhomogeneous one, and skin effect is strong, i.e. $\delta 
\ll a$ ($a$ is the characteristic size of the inhomogeneity), Eqs.(1) are local
up to the terms of the order of $\delta /a$. In other words, in Eqs.(1)
$\hat\zeta$ is an ordinary multiplying operator, but the elements of
$\hat\zeta$ depend on position at the surface. In this case Eqs.(1) are called
the local Leontovich (impedance) boundary conditions \cite{12,13}. Then to
solve an external electrodynamic problem it is necessary to know the surface
impedance tensor at every point of the metal surface. When the surface
inhomogeneity is strong, the solution requires cumbersome numerical
calculations.   

Sometimes it is sufficient to know the reflected electromagnetic wave averaged
over the surface inhomogeneities. The general way providing the correct result
is to derive equations for the averaged electromagnetic fields both in the
metal half-space and in the medium over the metal (we assume it to be vacuum),
as well as the boundary conditions for these fields at the metal-vacuum
interface. By proceeding from these equations we obtain the relation between
the tangential components of the averaged electric and magnetic vectors at the
metal surface and, consequently, the effective impedance tensor
${\hat\zeta}_{ef}$. Just this calculation procedure has been outlined in
Introduction.  
  
But when $\delta \ll a$ the problem is simplified because we can examine the 
fields in the vacuum side of the system only. Since in the vacuum Maxwell's 
equations do not contain any inhomogeneous parameters, the averaging of these 
equations gives the standard set of Maxwell's equations. Therefore the main 
problem is to obtain the boundary conditions for the averaged fields from the 
exact Eqs.(1). This can be done without the direct solution of the complete 
electrodynamic problem. 

The conception of the effective surface impedance is valid if the 
characteristic size of the surface inhomogeneity $a$ is small compared with 
the vacuum wave length $\lambda = 2\pi c/\omega$. Since the stochastic
fields are damped out at a distance of the order of $a$ from the metal surface,
beginning with a distances $d$, $a \ll d \ll \lambda$, the total
electromagnetic field equals to its averaged value. Then the problem is reduced
to the calculation of the electromagnetic fields reflected from the flat
surface whose surface impedance is ${\hat\zeta}_{ef}$. 

For polycrystals, the surface inhomogeneity is due to the misorientaion of the 
crystallographic axes of the grains at the metal surface, and the 
characteristic length $a$ is of the order of the mean size of a grain. In what 
follows we assume that
$$
\delta \ll a \ll \lambda.   \eqno (2)
$$

Let an electromagnetic wave be incident normally from the vacuum onto the
planar surface ${x}_{3} = 0$ of a polycrystalline metal which occupies the
region $x_3 < 0$. By analogy with Eq.(1), we define the effective surface
impedance tensor by the equation
$$
<{\bf E}_{t}> = {\hat \zeta}_{ef} [{\bf n},<{\bf H}_{t}>],    \eqno (3)
$$
where ${\bf n}$ coincides with ${\bf e}_{3}$; ${\bf e}_{3}$ is the unit vector
directed along the axis 3 of the laboratory coordinate system. The laboratory
coordinate system is related to the metal surface. The angular brackets denote
an average over the ensemble of realizations of the polycrystalline structure.

We assume that the polycrystalline medium we are concerned with is 
statistically homogeneous, i.e. that the ensemble averages are independent of
position. The only property of the medium that affects the one-point average,
for example, $<{\bf E}_{t}({\bf x})>$, is the rotation of the crystallographic
axes of the grain containing point $\bf x$ with respect to the laboratory
coordinate system. If the polycrystal is an untextured one, i.e. it is
composed of single crystal grains randomly rotated with respect to each other,
in Eq.(3) $<...>$ correspond to the averaging over all possible rotations of
the crystallographic axes of a grain at the surface.
 
To start the calculation, we would like to remind that since elements of the 
tensor $\hat\zeta$ are of the order of the ratio $\nu = \delta/\lambda \ll 1$,
the tangential components of the electric vector at the metal surface are
always much less than the tangential components of the magnetic vector. In the
lowest (the zeroth) order in $\nu$, the magnetic field strength at the metal
surface is equal to the magnetic vector ${\bf H}^{per}_{t}$ at the surface of a
perfect conductor. In the zeroth approximation in $\nu$ the vector ${\bf
E}_{t}$ is equal zero. The first nonvanishing term in the expansion of ${\bf
E}_{t}$ in powers of $\nu$ can be calculated with the aid of the local
Leontovich boundary conditions (1):   
$$
{\bf E}^{1}_{t} = \hat\zeta [{\bf n},{\bf H}^{per}_{t}].    \eqno (4)
$$

To obtain the averaged tangential components of the vectors $<{\bf H}_{t}>$ 
and $<{\bf E}_{t}>$ in the lowest orders in powers of $\nu$ (the zeroth and 
the first respectively) we need to know the vector ${\bf H}^{per}_{t}$ only:
$$
<{\bf H}_{t}>\, = \,<{\bf H}^{per}_{t}>;          \eqno (4.1)
$$
$$
<{\bf E}_{t}>\, = \,<{\bf E}_{t}^{1}>\, = \,
<\hat\zeta [{\bf n},{\bf H}^{per}_{t}]>.   
\eqno (4.2)
$$
If the vectors $<{\bf E}_{t}>$ and $<{\bf H}_{t}>$ are known, Eq.(3) defines 
the effective surface impedance ${\hat\zeta}_{ef}$.

Now it is almost obvious that when the surface of an inhomogeneous metal is 
flat, the effective surface impedance tensor $\hat\zeta$ is 
$$
{\hat\zeta}_{ef} = <\hat\zeta>.   \eqno (5)
$$
Indeed, in the case of normal incidence of an electromagnetic wave onto the 
flat surface ${x}_{3} = 0$ of a perfect conductor the magnetic vector at the 
surface is ${\bf H}_{t}^{per} = 2{\bf H}_{0}$, where ${\bf H}_{0}$ is the 
magnetic vector in the incident wave, and it does not depend on position at 
the surface. Next, the normal $\bf n$ to the surface ${x}_{3} = 0$ is also
independent of position. Thus, from Eqs.(4.1,2) we have:
$$
<{\bf H}_{t}> = 2{\bf H}_{0};          \eqno (6.1)
$$
$$
<{\bf E}_{t}> = 2<\hat\zeta>[{\bf n},{\bf H}_{0}].   \eqno (6.2)
$$
Comparing Eqs.(6) with Eq.(3) we obtain the above mentioned Eq.(5). 

In \cite{11} Eq.(5) was obtained according to the general scheme and the first 
correction ${\zeta}_{ef}^{1}$ to the effective impedance due to the local
impedance fluctuations was calculated. It has been shown that if the impedance
depends on position at the surface,
$$
{\zeta}_{\alpha\beta} = <\zeta>[{\delta}_{\alpha\beta} + 
{d}_{\alpha\beta}({\bf x}_{\parallel})]; \; 
<{d}_{\alpha\beta}({\bf x}_{\parallel})> = 0,   
$$
where ${\bf x}_{\parallel}$ is the two dimensional position vector at the
surface $x_3 = 0$, the correction ${\zeta}_{ef}^{1}$ is of the order of 
$$
{\zeta}_{ef}^{1} \sim \zeta\frac{\delta}{a}d^2 \sim 
\frac{{\delta}^{2}}{a\lambda}d^2;
$$
$d^2 = <{d}_{\alpha\beta}^{2}({\bf x}_{\parallel})>$. However, the Leontovich
boundary conditions (1) leads to the correct result for the amplitude of the
reflected wave only if $\delta /a \ll 1$. The terms of the order of 
${(\delta/a)}^{2}$ are outside of the framework of the Leontovich boundary
conditions applicability. Then the correction term ${\zeta}_{ef}^{1}$ has to
be omitted within the accuracy of the initial equations, Eqs.(1). Thus, Eq.(5)
gives the value of the effective surface impedance which can't be improved in
the framework of the Leontovich boundary conditions applicability. The result
is a nonperturbative one with respect to the inhomogeneity amplitude. It allows
to calculate the effective impedance of strongly anisotropic polycrystals.
\vskip 3mm
\section{EFFECTIVE IMPEDANCE OF POLYCRYSTAL UNDER  CONDITIONS OF 
ANOMALOUS SKIN EFFECT}

To make use of Eq.(5), we need to know the local impedance of a polycrystal 
under the conditions of anomalous skin effect ($\delta \ll l,\omega\tau \ll
1$). In this case the current density $\bf j$ and the electric field strength 
$\bf E$ are related by non-local integral equations:
$$
j_i = {\hat\sigma}_{ik}\{E_k\},   \eqno (7)
$$
where ${\hat\sigma}_{ik}$ is the matrix of linear integral operators; the 
braces $\{...\}$ denote that $\bf j$ is a functional of $\bf E$. To calculate
this operator the kinetic theory must be used \cite{16,17}.

For polycrystals the conductivity tensor ${\hat\sigma}_{ik}$ is a random 
operator. Since the crystallographic axes of the anisotropic single crystal 
grains are rotated randomly with respect to each other, the electromagnetic 
wave crossing the boundary between adjacent grains in effect moves from one 
anisotropic medium into a different one. 

Firstly, this means that when solving the kinetic equation, it must be taken 
into account that though the electron dispersion relation $\varepsilon = 
\varepsilon ({\bf p})$ is the same for all the grains, the Fermi surfaces of
the adjacent grains are rotated with respect to each other in 
the $\bf p$-space. Then the nonscalar parameters relating to the Fermi surface
written with respect to a fixed set of laboratory axes differ for different
grains. In particular, the electron velocity ${\bf v} = \partial \varepsilon
({\bf p})/\partial {\bf p}$ is not a function of the momentum $\bf p$ only,
but it depends on position $\bf x$ too: ${\bf v} = {\bf v}({\bf p};{\bf x})$.
(Naturally, the volume of the Fermi surface and the total area of its surface
are independent of $\bf x$.)  

Secondly, we need to specify the boundary conditions at the interfaces of the
grains. If the thickness of the grains interfaces is of the order of the atomic
spacing, the random functions are step functions across the interfaces. In the
model of the "spread" interfaces all the stochastic parameters across the
interfaces vary smoothly over the distances, which are much greater than the
atomic spacing, but much less than the mean size of the grain. Next, to
simplify the problem the relaxation time approximation (the
$\tau$-approximation) for the collision integral in the kinetic equation can be
used \cite{16,17}. It is known, that this approximation is sufficient
in the limiting case $l \gg \delta$. In our analysis we use the "spread"
interfaces model. Then the scattering at the grains interfaces has to be
included in the electron relaxation (mean free) time $\tau$, and $\tau$ itself
is a random function of position $\bf x$. 

Finally, boundary conditions at the metal surface $(x_3 = 0)$ and far from
the surface $(x_3 \to -\infty)$ have to be specified. It is evident that the 
calculation of the conductivity tensor for a polycrystal is a separate and 
a very difficult problem. 

To get a proper starting point for our calculation, we first of all assume that
the grains are sufficiently large and the mean size of the grains is much
larger than the electron mean free path\footnote{As it was mentioned in
Introduction $l \gg {\delta}_{p}$, where ${\delta}_{p} \sim
{10}^{-5}-{10}^{-6}{\rm cm}$ is the plasma penetration depth. Usually the mean
size of a polycrystalline grain $a \sim {10}^{-4}{\rm cm}$.} $l$,  
$$
a \gg l.   \eqno (8)
$$
Since $l$ is of the order of the distance over which the electric field $\bf E$
differs significantly, when Eq.(8) is fulfilled, the current density $\bf j$ in
a grain is nearly the same as in the single crystal rotated with
respect to the laboratory axes in the same way as the given particular grain. 

When calculating, we assume that the relaxation time $\tau$ is a constant. It
is not so essential, since usually when $\delta \ll l$ the relaxation time does
not enter the leading term of the expression for the surface impedance 
\cite{16,17}. In addition, as an excuse, we would like to note that if the
relaxation time depend on the electron energy only, $\tau = \tau
(\varepsilon)$, it does the same for all the grains. 
 
For single crystals the boundary conditions on the real surface of the metal
and the influence of these conditions on the impedance were analyzed by a lot
of authors (see, for example, ref. \cite{19}). It has been shown that under the
conditions of anomalous skin effect, the value of the surface impedance is not
very sensitive to the character of the reflection of electrons from the sample
surface. Therefore, we use the simplest boundary conditions and assume the
specular reflection of conductive electrons from the metal surface.

Under all these assumptions it is clear that, the local surface impedance at 
a point $\bf x$ on the surface of the polycrystal approximately equals to 
the surface impedance of the single crystal whose crystallographic axes are 
rotated with respect to the laboratory axes in the same way as the ones of the
grain at the point $\bf x$. Let's write down the elements of the local
impedance tensor. 

\subsection{The local surface impedance calculation}

As we have mentioned above, from the theory of skin effect in single crystal
metals it follows that under the conditions of extremely anomalous skin effect
the assumption of the specular reflection of conductive electrons from the
metal surface is sufficient. Then the Fourier method is efficient for the
surface impedance calculation \cite{16,17}. The elements of the surface
impedance tensor are expressed through their Fourier coefficients 
${\zeta}_{\alpha\beta}(k)$: 
$$
{\zeta}_{\alpha\beta} = \frac{1}{\pi}\int_0^\infty 
{\zeta}_{\alpha\beta}(k)dk.
\eqno (9)
$$
The two-dimensional tensor ${\zeta}_{\alpha\beta}(\alpha,\beta = 1,2)$ is
defined in the laboratory coordinate system related to the metal surface. The
axes $1$ and $2$ of this coordinate system are placed on the metal surface and
the axis $3$ is directed along the normal $\bf n$ to the surface. The Fourier
coefficients ${\zeta}_{\alpha\beta}(k)$ are expressed in terms of the elements
of the tensor ${\hat\zeta}^{-1}(k)$ reciprocal to ${\zeta}_{\alpha\beta}(k)$:
$$
{{\zeta}_{\alpha\beta}}^{-1}(k) = -\frac{c}{2i\omega}
[{k}^{2}{\delta}_{\alpha\beta} - \frac{4\pi i\omega}{{c}^{2}}
{\sigma}_{\alpha\beta}(k)].     \eqno (10)
$$
The explicit form of the elements of the tensor ${\zeta}_{\alpha\beta}(k)$ is:
$$
{\zeta}_{11}(k) = {{\zeta}_{22}}^{-1}(k)/Z(k);\;   
{\zeta}_{12}(k) = - {{\zeta}_{12}}^{-1}(k)/Z(k);\;   
{\zeta}_{22}(k) = {{\zeta}_{11}}^{-1}(k)/Z(k),   \eqno (11)
$$
where $Z(k) = {{\zeta}_{11}}^{-1}(k){{\zeta}_{22}}^{-1}(k) - 
{[{{\zeta}_{12}}^{-1}(k)]}^{2}$, or
$$
Z(k) = -{\left(\frac{c}{2\omega}\right)}^{2}
\left\{ k^4 - \frac{4\pi i\omega}{c^2} k^2 ({\sigma}_{11}(k) + 
{\sigma}_{22}(k))
+ {\left(\frac{4\pi i\omega}{c^2}\right)}^{2}[{\sigma}_{11}(k)
{\sigma}_{22}(k) - {\sigma}^{2}_{12}(k)]\right\}.
\eqno (12)
$$

In equations (10) and (12) ${\sigma}_{ik}(k)$ are the Fourier coefficients of
the elements of the conductivity tensor calculated for the unbounded single
crystal metal. In the $\tau$-approximation \cite{16,17}
$$
{\sigma}_{ik}({\bf k}) = \frac{2e^2 \tau}{{(2\pi\hbar)}^{3}}
\int \frac{{v}_{i}{v}_{k}}{v[1+i{\bf kv}\tau]} dS,
\eqno (13)
$$
where the integration is carried out over the Fermi surface in a single cell of
the momentum space; $v = |{\bf v}({\bf p})|$. The Fermi surface is defined by
the equation $\varepsilon ({\bf p}) = {\varepsilon}_{F}$; ${\varepsilon}_{F}$
is the Fermi energy. Taking into account that every Fermi surface has an
inversion center, Eq.(13) can be rewritten as
$$
{\sigma}_{ik}({\bf k}) = \frac{4e^2 \tau}{{(2\pi\hbar)}^{3}}
\int \frac{{v}_{i}{v}_{k}}{v[1+{({\bf kv}\tau)}^{2}]} dS;  \eqno (13.1)
$$
the integration is carried out over part of the Fermi surface where 
${\bf kv} > 0$. The last equation shows that the elements of the tensor
${\sigma}_{ik}({\bf k})$ are real. When calculating the surface impedance the
wave vector $\bf k$ is supposed to be directed along the normale to the metal
surface, i.e. with respect to the laboratory coordinate system 
${\bf k} = (0,0,k)$.  

To avoid awkwardness, in Eqs.(10) - (12) we do not indicate explicitly the 
dependence of the elements of the conductivity tensor on the direction of the
wave vector $\bf k$ with respect to the orientation of the crystallographic
axes, but since we are interested just in the orientational dependence of the
surface impedance, it is necessary to have this dependence in mind. 

In what follows we use Eq.(13) under the conditions of extremely anomalous skin
effect. The integral (nonlocal) relation between the electric field and the
current is manifested in the dependence of the conductivity tensor
${\sigma}_{ik}$ on the wave vector $\bf k$ when $kl \sim l/\delta \gg 1$;
$\delta$ is the characteristic penetration depth of electromagnetic field into
the metal and $l \sim v\tau$. If $kl\gg 1$, usually only the transverse with
respect to the vector $\bf k$ elements of the conductivity tensor contribute to
the current density. They are the functionals of the Fermi surface Gaussian
curvature at the points where ${\bf kv}=0$ (see refs. \cite{16,17}). 

It is of interest that in the limit $kl \gg 1$ in spite of such complicated
dependence of ${\sigma}_{ik}(\bf k)$ on the Fermi surface geometry, for a given
vector $\bf k$ the conductivity tensor averaged over all orientations of
the crystallographic axes is given by a very simple expression
\cite{8}: 
$$
<{\sigma}_{ik}({\bf k})> = {\sigma}_{a}({\delta}_{ik} - 
{k}_{i}{k}_{k}/{k}^{2}), \eqno (14.1)
$$
$$
{\sigma}_{a} = \frac{\pi {e}^{2}{S}_{F}}{2{(2\pi \hbar)}^{3}k},
\eqno (14.2)
$$
${S}_{F}$ is the total area of the Fermi surface. We see that, as it must be,
the dependence of $<{\sigma}_{ik}({\bf k})>$ on the direction of the vector
$\bf k$ is the same as for an isotropic metal.  

For slightly anisotropic polycrystals under the conditions of extremely
anomalous skin-effect Eqs.(14) define the surface impedance of the polycrystal
in the zeroth approximation with respect to anisotropy. Apparently, the small
anisotropy means either that the Fermi surface is regularly close to a sphere
(for example, it is an ellipsoid with nearly equal principal axes), or the
"weight" of the regions where the Fermi surface deviates from the sphere is
small. It has been shown in ref. \cite{11} 
that with regard to the effective surface impedance calculation the Fermi
surface anisotropy can be considered small, if 
$$
{\Delta}^{2}  = \frac{4}{\pi {S}^{2}_{F}}\int \int 
\frac{dS_1 dS_2 {({\vec\nu}_{1}{\vec\nu}_{2})}^{2}}
{\sqrt{1-{({\vec\nu}_{1}{\vec\nu}_{2})}^{2}}}\; -1 \; \ll \; 1.
\eqno (15)
$$
In Eq.(15) the double integration is carried out over the Fermi surface;
$\vec \nu = {\bf v}({\bf p})/v({\bf p})$ is the unit vector normal to the
Fermi surface in the point $\bf p$.

If inequality (15) is fulfilled, we can replace the true Fermi surface by a
sphere whose surface area is $S_F$. The surface impedance of such an isotropic
conductor  is ${\zeta}_{\alpha\beta} = {\zeta}_{a}{\delta}_{\alpha\beta}$,
where
$$
{\zeta}_{a} = \frac{2(1-i\sqrt{3})}{3\sqrt{3}}
\left(\frac{\omega{\delta}_{a}}{c}\right),\quad
{\delta}_{a} = {\left(\frac{4\pi{c}^{2}{\hbar}^{3}}
{\omega{e}^{2}{S}_{F}}\right)}^{1/3};
\eqno (16)
$$
${\delta}_{a}$ is the relevant electric field penetration depth \cite{16}.
Equation (16) gives the leading term of the expression for the effective
impedance. The first correction to ${\zeta}_{a}$ is proportional to 
${\Delta}^{2}$ (see ref. \cite{11}). 

When anisotropy is strong, equations (14) and (16) are inapplicable. In what
follows, we use Eq.(5) to calculate the effective impedance of polycrystals
with an arbitrary anisotropy. The difference between ${\zeta}_{ef}$ and
${\zeta}_{a}$ shows the dependence of the effective impedance on the Fermi
surface geometry and exhibits the difference between the polycrystal and an
isotropic conductor with the spherical Fermi surface. 

Here we would like to mention that strictly speaking the averaged conductivity
(14.1) has no specific physical meaning. We use ${\sigma}_{a}$ defined by
Eq.(14.2) only as a characteristic value of the conductivity relating to the
given Fermi surface. In just the same way we use ${\zeta}_{a}$ and
${\delta}_{a}$ (see Eq.(16)) as the characteristic values of the surface
impedance and the penetration depth respectively.  

To calculate ${\zeta}_{ef}$ we need to know the elements of the local impedance
tensor ${\zeta}_{\alpha\beta}$. With regard to the arguments presented above,
we use Eq.(9). The Fourier coefficients ${\zeta}_{\alpha\beta}(k)$ in Eq.(9)
are expressed in terms of the elements of the conductivity tensor
${\sigma}_{ik}(\bf k)$ (see Eq.(13)). Thus, the first step of our calculation
is to obtain the elements of this tensor for an arbitrary orientation of the
wave vector $\bf k$ with respect to the crystallographic axes. (We remind:
${\bf k} \parallel {\bf n}$.) 

As a rule, under the conditions of extremely anomalous skin effect ($kl \gg 1$)
when calculating the elements of the conductivity tensor the leading terms in
the expressions for the elements of ${\sigma}_{ik}(\bf k)$ can be written in
the form   
$$
{\sigma}_{ik}({\bf k}) = {\sigma}_{a}(k){S}_{ik}; \eqno (17.1)   
$$
where ${\sigma}_{a}$ is given by Eq.(14.2) and the elements of the
dimensionless tensor ${S}_{ik}$ depend on the Fermi surface geometry and the
orientation of the vector $\bf k$ only: 
$$
{S}_{ik} = \frac{4}{{S}_{F}}\int {\nu}_{i}{\nu}_{k} \delta ({\bf kv}/kv)dS, 
 \eqno (17.2)
$$
${\nu}_{i} = {v}_{i}({\bf p})/v({\bf p})$ and $\delta (x)$ is the 
delta-function. 

Usually it is most covenient to calculate the elements of the conductivity
tensor with respect to the crystallographic axes. Let us introduce two sets of
coordinate axes: a fixed set of laboratory axes related to the surface of the
polycrystal and the crystallographic axes, which are rotated with respect to
the laboratory axes. Let ${\bf e}_{i}$ be the unit vectors along the laboratory
axes (${\bf e}_{3} \parallel {\bf n}$) and ${\bf a}_{i}$ be the unit vectors
along the crystallographic axes. Let $\gamma$ denote the rotation of the
crystallographic axes through the Euler angles
${\theta}_{k},{\psi}_{k},{\varphi}_{k}$. We introduce the rotation matrix   
$$
{\alpha}_{ik} (\gamma) = ({\bf e}_{i}{\bf a}_{k}).   \eqno (18)
$$
The explicit form of the elements of the rotation matrix (18) is presented in
Appendix 1. 

Since the wave vector $\bf k$ written with respect to the laboratory coordinate
system is ${\bf k} = (0,0,k)$, its components written with respect to the
crystallographic axes depend on $\gamma$: ${k}_{i}^{(a)}(\gamma) = 
k{\alpha}_{3i}$. (In what follows the superscript $(a)$ denotes vectors and
tensors written with respect to crystallographic coordinate system.) Then
according to Eqs.(17) the elements of the tensor ${\sigma}_{ik}^{(a)}({\bf k})$
are functions of the Euler angles too. Finally, the elements of the
conductivity tensor with respect to the laboratory axes are\footnote{We do not
use special superscripts to indicate the laboratory coordinate system. Up to
the end of this Section, vectors and tensors without superscripts are written
with respect to the laboratory coordinate system.}:   
$$
{\sigma}_{ik}(k;\gamma) = {\sigma}_{a}(k){S}_{ik}(\gamma );\qquad
{S}_{ik}(\gamma ) = {\alpha}_{ip}(\gamma){\alpha}_{kq}(\gamma)
{S}_{pq}^{(a)}(\gamma). \eqno (19.1)
$$
From Eqs.(17) it follows that usually the tangential elements of the tensor
${S}_{ik}(\gamma )$ are of the order of unity and  
$$
{\sigma}_{\alpha\beta}(k;\gamma) = {\sigma}_{a}(k){S}_{\alpha\beta}(\gamma );
\qquad \alpha, \beta = 1,2. \eqno (19.2)
$$
In the same approximation the elements ${S}_{i3}(\gamma ) = 0;i = 1,2,3$. When
in Eqs.(13) the next terms in the small parameter $1/kl$ are taken into
account, it can be shown that    
$$
{\sigma}_{13} \approx {\sigma}_{23} \approx {\sigma}_{33} \sim 
\frac{1}{kl}{\sigma}_{a},    \eqno (19.3)
$$
If by any reason the tangential elements ${S}_{\alpha\beta}$ of the tensor 
$\hat S$ are equal to zero, all the elements of the tensor ${\sigma}_{ik}(k)$
are of the order of ${\sigma}_{a}/kl$. In Section IV we examine several
examples of such extraordinary situations, but here we restrict ourselves to
Eqs.(19.2).   

Now we can calculate the elements of the local impedance tensor. We rewrite
Eqs.(10)-(12) for the Fourier coefficients ${\zeta}_{\alpha\beta}(k)$ in terms
of the dimensionless tensor ${S}_{\alpha\beta}(\gamma)$. Then
$$
Z(k;\gamma) = -\frac{1}{{\delta}^{4}_{a} {k}^{2}}
{\left(\frac{c}{2\omega}\right)}^{2}z(x;\gamma),  \eqno (20.1)
$$
where $x = k{\delta}_{a}$, and ${\delta}_{a}$ is given by Eq.(16);
$$
z(x;\gamma) = {x}^{6} - i{x}^{3}({S}_{11}(\gamma)+{S}_{22}(\gamma)) - 
[{S}_{11}(\gamma){S}_{22}(\gamma)-{S}_{12}^2 (\gamma)].   \eqno (20.2) 
$$
And the Fourier coefficients of the elements of the surface impedance tensor
are 
$$
{\zeta}_{11}(x;\gamma ) = -i\frac{2\omega{\delta}_{a}}{c}{\delta}_{a}x
[{x}^{3} -i{S}_{22}(\gamma)]/z(x;\gamma);   \eqno (21.1)
$$
$$
{\zeta}_{12}(x;\gamma ) = \frac{2\omega{\delta}_{a}}{c}{\delta}_{a}x 
{S}_{12}(\gamma)/z(x;\gamma); 
\eqno (21.2)
$$
$$
{\zeta}_{22}(x;\gamma) = -i\frac{2\omega{\delta}_{a}}{c}{\delta }_{a}x
[{x}^{3} -i{S}_{11}(\gamma)]/z(x;\gamma),   \eqno (21.3)
$$

We substitute Eqs.(21) into Eq.(9). To carry out the integration, we rewrite
Eq.(20.2) for the function $z(x;\gamma)$ in the form
$$
z(x;\gamma) = (x^3 -iS_1(\gamma ))(x^3 -iS_2(\gamma )),  \eqno (22.1)
$$
where the functions $S_1(\gamma )$ and $S_2(\gamma )$ are the principal values
of the two-dimensional tensor ${S}_{\alpha\beta}(\gamma)$:
$$
{S}_{1,2}(\gamma) = \frac{1}{2}[({S}_{11}(\gamma) + {S}_{22}(\gamma) \pm
R(\gamma)],\; 
R(\gamma) = \sqrt{{({S}_{11}(\gamma)-{S}_{22}(\gamma))}^{2} +
4{S}_{12}^{2}(\gamma)}.   \eqno (22.2)
$$
Equation (22.1) defines the poles of the integrand in the expressions (9) for
the elements of the local impedance tensor ${\zeta}_{\alpha\beta}(\gamma)$ (The
method of calculation of the integrals is given, for example, in ref.
\cite{16}.) 

After the integration is carried out, we obtain the elements of the local
impedance tensor as functions of $\gamma$:  
$$
{\zeta}_{11}(\gamma) = \frac{1}{2}{\zeta}_{a}\{({{S}_{1}}^{-1/3}(\gamma) + 
{{S}_{2}}^{-1/3}(\gamma)) + s(\gamma)({{S}_{1}}^{-1/3}(\gamma) -
{{S}_{2}}^{-1/3}(\gamma))\},   \eqno (23.1)
$$ 
$$
{\zeta}_{22}(\gamma) = \frac{1}{2}{\zeta}_{a}\{({{S}_{1}}^{-1/3}(\gamma) + 
{{S}_{2}}^{-1/3}(\gamma)) - s(\gamma)({{S}_{1}}^{-1/3}(\gamma) -
{{S}_{2}}^{-1/3}(\gamma))\},   \eqno (23.2)
$$
with ${\zeta}_{a}$ from Eq.(16) and
$$
s(\gamma) = \frac{({S}_{11}(\gamma)-{S}_{22}(\gamma))}{R(\gamma)}. \eqno (23.3)
$$
In Eqs.(20)-(23) the dependence of all the terms on the Euler angles (on the 
set $\gamma$) is shown  explicitly. We do not write down the expression for
${\zeta}_{12}$, since that element of the surface impedance tensor does not
contribute to ${\zeta}_{ef}$. 

\subsection{The effective surface impedance calculation}

Now we are ready for the second step of our calculation. In accordance with
Eq.(5) the elements of the effective surface impedance tensor are the averages
over the rotations $\gamma$ of the local impedance tensor (23). The averaging
includes the integration over all of the three Euler angles
${\theta}_{k},{\psi}_{k},{\varphi}_{k}$ (see Eq.(A1.2)). With regard to our
definition of the Euler angles (see Appendix 1) the direct calculation shows
that in equations (23.1) and (23.2) the Euler angle ${\varphi}_{k}$ enters only
the expression for the function $s(\gamma)$. The structure of this function is:
$s(\gamma) = S({\theta}_{k},{\psi}_{k})\sin 2{\varphi}_{k} + 
C({\theta}_{k},{\psi}_{k})\cos 2{\varphi}_{k}$. We also showed that the
nondiagonal element ${S}_{12}$ of the tensor $\hat S(\gamma )$ depended on the
angle ${\varphi}_{k}$ in the same way as the function $s(\gamma )$. (In
Appendix 2 we present the elements of the tensor $\hat S(\gamma )$ calculated
for an axially symmetric Fermi surface.) Then, after the integration over the
angle ${\varphi}_{k}$, it is evident that 
$$
{\hat\zeta}_{ef} = {\zeta}_{ef}\hat I,   \eqno (24.1)
$$
$\hat I$ is the two-dimensional unit matrix, and
$$
{\zeta}_{ef} = \frac{1}{2}{\zeta}_{a}<{S}_{1}^{-1/3}({\theta}_{k},{\psi}_{k}) +
{S}_{2}^{-1/3}({\theta}_{k},{\psi}_{k})>,  \eqno (24.2)
$$
where
$$
<...> = \frac{1}{4\pi}\int_0^{\pi} \sin {\theta}_{k}d{\theta}_{k}
\int_0^{2\pi} ... d{\psi}_{k}.  \eqno (24.3)
$$
With the aid of Eqs.(24), the effective surface impedance of a polycrystalline
metal can be calculated (at least numerically), if the equation of the Fermi
surface of the original single crystal is known. 

Here we would like to note, that according to Eq.(24.2) if the average
$<{{S}_{1}}^{-1/3} + {{S}_{2}}^{-1/3}>$ is not extremely large or small,
${\zeta}_{ef}$ in the order of magnitude is the same as ${\zeta}_{a}$. We
remind, that ${\zeta}_{a}$ is the surface impedance of an isotropic conductor
with the spherical Fermi surface, whose area is equal ${S}_{F}$. Then the
difference between ${\zeta}_{ef}$ and ${\zeta}_{a}$ is in a numerical factor
which can be calculated with regard to Eqs.(24). Next, since in Eq.(24.2) the
functions ${S}_{1(2)}({\theta}_{k},{\psi}_{k})$ are real, the relation between
the real and the imaginary parts of ${\zeta}_{ef}$ is given by the factor 
$(1-i\sqrt{3})$ in the expression for ${\zeta}_{a}$ (see Eq.(16)). Under the
conditions of anomalous skin effect the same factor appears usually in the
expressions for the elements of the impedance tensor \cite{16}. It is defined
by the poles of the type $k^3 - iC$ in the expressions for the Fourier
coefficients of the elements of the impedance tensor (see Eqs. (21) and (22)).
Several examples of the Fermi surfaces for which this general rule is not true
are presented below in Subsections IV.B and IV.C.  

Equations (24) are rather formal. In what follows we present the formulae
obtained from Eqs.(24) for a Fermi surface which is a surface of revolution.
The derivation of Eqs.(25) is presented in Appendix 2.

Let the rotation axis of the Fermi surface coincide with the crystallographic 
axis $z$. (For axially symmetric Fermi surfaces by axis $z$ we mean the
rotation axis and use a subscript $\bot$ for the vectors in the plane
perpendicular to the axis $z$.) Let the equation of the Fermi surface written
with respect to the crystallographic axes be ${\varepsilon}_{F}= 
\varepsilon ({\bf p}) = \varepsilon({p}_{\bot},{p}_{z})$, ${p}_{\bot} = 
|{\bf p}_{\bot}|$. We introduce the transverse speed of an electron on the
Fermi surface ${v}_{\bot}= \partial \varepsilon ({p}_{\bot},{p}_{z})/\partial
{p}_{\bot}$ and the projection of an electron velocity on the axis $z$:
${v}_{z}= \partial \varepsilon ({p}_{\bot},{p}_{z})/\partial {p}_{z}$.  

As it is shown in Appendix 2, in this case the functions ${S}_{1,2}$ (see
Eqs.(22)), depend not on the three Euler angles composing the rotation 
$\gamma$, but on the spherical angle ${\theta}_{k}$ only:
$$
{S}_{1} = \frac{16}{{S}_{F}\sin {\theta}_{k}\tan {\theta}_{k}}
\int {p}_{\bot} \Phi ({p}_{\bot}, \tan {\theta}_{k}) d{p}_{\bot}; \eqno (25.1)
$$
$$
{S}_{2} = \frac{16\tan {\theta}_{k}}{{S}_{F}\sin^3 {\theta}_{k}}
\int {p}_{\bot} \frac {d{p}_{\bot}}{\Phi ({p}_{\bot}, \tan {\theta}_{k})},
\eqno (25.2)
$$
where
$$
\Phi ({p}_{\bot}, \tan {\theta}_{k}) = \sqrt
{\frac{{v}_{\bot}^{2}\tan^2 {\theta}_{k} -{v}_{z}^{2} }{{v}_{z}^{2}}}.
\eqno (25.3)
$$
We do not use the superscript $(a)$ here, but we have in mind that ${p}_{\bot},
{p}_{z}$ and ${v}_{\bot}, {v}_{z}$ are calculated with respect to the 
crystallographic coordinate system. In Eqs.(25) the integration is carried out
over the part of the Fermi surface inside one cell of the momentum space, where
$v_z > 0$ and the radicand of the function $\Phi ({p}_{\bot}, \tan
{\theta}_{k})$ is positive. We would like to mention, that if the Fermi surface
is a multiply connected surface, the integration is spread over all of its
parts. 

Next, when the Fermi surface is an axially symmetric surface the function 
$s(\gamma) = -\cos 2{\varphi}_{k}$ (see Eq.(A2.6b)). Consequently, in
accordance with the general Eqs.(24) we have: 
$$
{\zeta}_{ef}{\delta}_{\alpha\beta}
= \frac{1}{4\pi}\int_0^{\pi}\sin {\theta}_{k}{\rm d}{\theta}_{k}
\int_0^{2\pi}{\rm d}{\varphi}_{k}
{\zeta}_{\alpha\beta}({\theta}_{k},{\varphi}_{k}).   \eqno (26)
$$
And after integration over ${\varphi}_{k}$ our result is:
$$
{\zeta}_{ef} = \frac{1}{2}{\zeta}_{a}<{{S}_{1}}^{-1/3} + {{S}_{2}}^{-1/3}>,
\eqno (27.1)
$$
where
$$
<{{S}_{\alpha}}^{-1/3}> = \int_0^{\pi/2}\sin {\theta}_{k}
{{S}_{\alpha}}^{-1/3}{\rm d}{\theta}_{k}; \quad \alpha = 1,2.    \eqno (27.2)
$$
It is easy to see, that for an isotropic metal with a spherical Fermi surface
${S}_{1}={S}_{2}=1$ and ${\zeta}_{ef}={\zeta}_{a}$.

In conclusion of this Section we would like to point out that the obtained 
Eqs.(27) are rather simple. Under the conditions of anomalous skin effect
they allow us to calculate the effective surface impedance of a polycrystal if
the dispersion relation of the conduction electrons (in other words, the
equation of the Fermi surface) is known for the original single crystal metal.
Moreover, Eqs.(27) allow to investigate the influence of the Fermi surface
geometry on the value of the effective surface impedance.
\vskip 3mm

\section{EFFECTIVE SURFACE IMPEDANCE OF POLYCRYSTALS COMPOSED OF THE 
GRAINS WITH SOME MODEL FERMI SURFACES}
 
In this Section, we present some examples of the effective surface impedance
calculation for different model polycrystals. We assume that Fermi surfaces of
original single crystals have rather simple forms. Although the examples 
discussed below cannot be directly related to real metals, they allow us to
solve the problem accurately (up to the numerical factors) and to show clearly
the dependence of the effective impedance on the geometry of the Fermi surface.

\subsection{An ellipsoidal Fermi surface}

Let the Fermi surface be an uniaxial ellipsoid. Such a surface is the simplest
example of a closed nonspherical Fermi surface. With respect to the
crystallographic axes, the equation of the Fermi surface is
$$
{\varepsilon}_{F} = \frac{1}{2{m}_{\bot}}{p}_{\bot}^{2} +
\frac{1}{2{m}_{z}}{p}_{z}^{2};  \eqno (28)
$$
We introduce
$$
{p}_{*} = {(2{m}_{\bot}{\varepsilon}_{F})}^{1/2};\qquad 
\mu = {m}_{z}/{m}_{\bot}. \eqno (29)
$$ 
If $\mu \ll 1$, the Fermi surface is close to a disk; if $\mu \gg 1$, it is a
needle-shaped one. If $\mu = 1$, Eq.(28) corresponds to an isotropic conductor
with spherical Fermi surface. The total area ${S}_{F}$ of the surface 
(28) in terms of ${p}_{*}$ and $\mu$ is:
$$
{S}_{F} = 4\pi {p}_{*}^{2}Q(\mu), 
$$
$$
Q(\mu ) = \frac{1}{2}\left\{1+ \frac{\mu}{2\sqrt{1-\mu}}
\ln \left[\frac{1+\sqrt{1-\mu}}{1-\sqrt{1-\mu}}\right]\right\},
\qquad {\rm if}\quad \mu < 1;  \eqno (30)
$$
$$
Q(\mu ) = \frac{1}{2}\left\{1+\frac{\mu}{\sqrt{\mu -1}}
\arcsin \sqrt{\frac{\mu -1}{\mu}}\right\},
\qquad {\rm if}\quad \mu >1.   
$$

With regard to Eq.(27.1) to calculate the effective impedance we have to
calculate the functions ${S}_{1,2}({\theta}_{k})$ defined by
Eqs.(25)\footnote{We remind that ${S}_{1,2}$ are the principal values of the
tensor ${S}_{\alpha\beta}$.}. Since the radicand of the function  $\Phi
({p}_{\bot}, \tan {\theta}_{k})$ (see Eq.(25.3)) is positive when  
${p}_{*}/\sqrt{1+\mu\tan^2 {\theta}_{k}} < {p}_{\bot} < {p}_{*}$, just this
interval is the domain of integration in equations (25.1) and (25.2). Then  
$$
{S}_{1}({\theta}_{k};\mu ) = \frac{\mu }{Q(\mu )\sqrt{\cos^2 {\theta}_{k}
+ \mu \sin^2 {\theta}_{k}}}; \qquad
{S}_{2}({\theta}_{k};\mu ) = \frac{\mu }{Q(\mu ){[\cos^2 {\theta}_{k}
+ \mu \sin^2 {\theta}_{k}]}^{3/2}}. \eqno (31.1)
$$
For the following, let's write down ${S}_{1,2}$ for nearly disk-shaped 
($\mu \ll 1$) and needle-shaped ($\mu \gg 1$) ellipsoids, i.e in the cases of
strongly anisotropic Fermi surfaces (28). From equations (30) and (31.1) we
obtain:   
$$
{S}_{1} \approx \frac{2\mu}{|\cos {\theta}_{k}|}; \quad
{S}_{2} \approx \frac{2\mu}{|\cos^3 {\theta}_{k}|}, \quad {\rm when}\quad
 \mu \ll 1. \eqno (31.2)
$$
$$
{S}_{1} \approx \frac{4}{\pi \sin {\theta}_{k}}; \quad
{S}_{2} \approx \frac{4}{\pi\mu \sin^3 {\theta}_{k}}, \quad {\rm when} \quad
\mu \gg 1.  \eqno (31.3)
$$
Thus, in these limiting cases at least one of the principle values of the
tensor ${S}_{\alpha\beta}$ is singularly small. In other words, at least one of
the principle conductivities is singularly small compared with the averaged
conductivity ${\sigma}_{a}$ (see Eq.(19.2)).    

Now with regard to Eqs.(27), the effective impedance is
$$
{\zeta}_{ef}^{(el)} = {\zeta}_{a}Z(\mu ),  \eqno (32.1)
$$
where ${\zeta}_{a}$ is defined by Eq.(16) and
$$
Z(\mu ) = \frac{1}{2}{\left(\frac{Q(\mu )}{\mu}\right)}^{1/3}
\int_0^1 dx {M}^{1/6}(x;\mu)[1+{M}^{1/3}(x;\mu)];\quad
M(x;\mu ) = x^2 +\mu (1-x^2).  \eqno (32.2)
$$
The function $Z(\mu )$ is presented in Fig.1. It can be shown, that 
$$
Z(\mu ) \approx \frac{5}{8}{\left(\frac{1}{2\mu}\right)}^{1/3},\;  {\rm if} 
\quad \mu \ll 1 \qquad {\rm and} \qquad Z (\mu )\approx \frac{\pi}{8}
{\left(\frac{\pi \mu}{4}\right)}^{1/3},\; {\rm if}\quad \mu \gg 1. \eqno (32.3)
$$
When the anisotropy is strong, the function $Z(\mu) \gg 1$. Consequently,
in these cases the effective impedance ${\zeta}_{ef}^{(el)} \gg {\zeta}_{a}$.

We can rewrite Eq.(32.1) for ${\zeta}_{ef}$ in the form similar to Eq.(16) for
${\zeta}_{a}$ introducing an effective area of the Fermi surface
${S}_{ef}$. Comparing equations (16) and (32.1) we obtain:
$$
{\zeta}_{ef} = \frac{2(1-i\sqrt{3})}{3\sqrt{3}}
\left(\frac{\omega{\delta}_{ef}}{c}\right),\quad
{\delta}_{ef} = {\left(\frac{4\pi{c}^{2}{\hbar}^{3}}
{\omega{e}^{2}{S}_{ef}}\right)}^{1/3},\quad
{S}_{ef} = S_F{Z}^{-3}(\mu).    \eqno (33.1)
$$
For strongly anisotropic Fermi surfaces (28) the effective area ${S}_{ef}$ is
much less than ${S}_{F}$. In terms of the effective masses ${m}_{\bot}$ and
$m_z$ we have 
$$
{S}_{ef} \sim {\varepsilon}_{F}m_z;\qquad {\rm if}
\quad {m}_{z} \ll {m}_{\bot} \quad ({\rm or}\quad \mu \ll 1); \eqno (33.2) 
$$
$$
{S}_{ef} \sim {\varepsilon}_{F}{m}_{\bot}\sqrt{\frac{{m}_{\bot}}{m_z}};
\qquad {\rm if}\quad {m}_{z} \gg {m}_{\bot} \quad ({\rm or}\quad \mu \gg 1); 
\eqno (33.3)
$$
The last two equations are in agreement with the results of \cite{10,11} for
strongly flattened and strongly elongated ellipsoids.  

We can also present our result, Eqs.(32), in a form showing the dependence of
the effective impedance on the electron number density $n$. Since 
$n = 2{V}_{F}/{(2\pi\hbar)}^{3}$, where ${V}_{F}$ is the volume of the Fermi
surface, we can express ${S}_{F}$ in terms of $n$ and $\mu$. Then, with regard
to Eqs.(33.1) we have 
$$
{\zeta}_{ef} = \frac{2(1-i\sqrt{3})}{3\sqrt{3}}
{\left(\frac{{\omega}^{2}\hbar}{e^2 c}\right)}^{1/3} 
{\left(\frac{1}{3{\pi}^{2}n}\right)}^{2/9}\tilde Z(\mu);\quad
\tilde Z(\mu) = \frac{{\mu}^{1/9}}{{Q(\mu)}^{1/3}}Z(\mu ).  \eqno (34.1)
$$
If the anisotropy is strong, with regard to the definition (30) the function 
$\tilde Z(\mu)$ is
$$
\tilde Z(\mu ) \approx \frac{5}{8}{\mu}^{-2/9},\;  {\rm if} 
\quad \mu \ll 1 \qquad {\rm and} \qquad \tilde Z (\mu )\approx \frac{\pi}{8}
{\mu}^{5/18},\; {\rm if}\quad \mu \gg 1. \eqno (34.2)
$$

Our calculation shows, that although usually ${\zeta}_{a}$ can be used as an
estimate of the effective impedance of a polycrystal, there are situations when
${\zeta}_{ef}$  differs from ${\zeta}_{a}$ significantly. They are those
"extraordinary situations" which were mentioned in Section III when discussing
the dependence of the elements of the conductivity tensor ${\sigma}_{ik}
({\bf k})$ on the value of $kl$. Let's clarify the situation.  

Let's assume that the Fermi surface is a thin disk. Then everywhere at the
surface the electron velocity ${\bf v} = (0,0,v)$, and the condition ${\bf
kv}=0$ can be fulfilled only if the wave vector $\bf k$ is situated at the
plane of the disk (${\theta}_{k}=\pi /2$). In other words, for an arbitrary
directed vector $\bf k$ there are no "belts" on the disk-shaped Fermi surface
providing the terms of the order of ${\sigma}_{a}$ in the series expansion of
the elements of the conductivity tensor ${\sigma}_{ik}(k)$ in powers of the
small parameter $1/kl$ (see Eqs.(17)). In the case of the disk-shaped Fermi
surface, the series expansions of all the elements of the conductivity tensor
${\sigma}_{ik}({\bf k})$ begin with the terms of the order of ${\sigma}_{a}/kl
\ll {\sigma}_{a}$. It is known, the less the conductivity, the greater the
impedance. Thus, the effective impedance of a polycrystal composed of the
grains with the disk-shaped Fermi surface has to be much greater than
${\zeta}_{a}$.   

When the Fermi surface is not a thin disk but a strongly flattened ellipsoid
($\mu \ll 1$), for an arbitrary direction of the wave vector $\bf k$ there are
"the belts" on the Fermi surface. For an ellipsoid the equation of "the belt"
is   
$$
{p}_{\bot} = \frac{{p}_{*}}{\sqrt{1+\mu\cos^2 \varphi\tan^2 {\theta}_{k}}},
$$
where $\varphi$ is the polar angle in the plane perpendicular to the rotation
axis and ${p}_{*}/\sqrt{1+\mu \tan^2 {\theta}_{k}} < {p}_{\bot} < {p}_{*}$.
If $\mu \ll 1$, almost all "the belts" are placed near the vertexes of the
ellipsoid related to its major diameter. However, "the belts" are very small:
along "the belts" the Gaussian curvature of the Fermi surface\footnote{Usually
in textbooks of electron theory of metals the elements of the conductivity
tensor ${\sigma}_{\alpha\beta}(k)$ are written as functionals of the Gaussian
curvature of the Fermi surface along "the belt" ${\bf kv} = 0$ (see, for
example, ref.\cite{16}). The greater the curvature of the belt, the less the
conductivity.} is much greater than $1/{p}_{*}^{2}$. This is the reason why
both principle values of the tensor ${S}_{\alpha\beta}$ are much less than
unity (see Eq.(31.2)) and, consequently, the effective impedance is much
greater than ${\zeta}_{a}$.   

The case of a strongly elongated ellipsoid ($\mu \gg 1$) is alike, but not
exactly the same as the one discussed above. Here almost everywhere at the
Fermi surface ${v}_{\bot} \gg v_z$. When the angle ${\theta}_{k}$ is not very
small ($\tan {\theta}_{k} \gg 1/\sqrt{\mu}$), "the belts" are placed mainly
near the vertexes of the ellipsoid related to the rotation axis $z$. Again,
for almost all the angles $\varphi$, the Gaussian curvature of the Fermi
surface along "the belts" is much greater than $1/{p}_{*}^{2}$. As a result one
of the principle values of the tensor ${S}_{ik}$, namely ${S}_{2}$ (see 
Eq.(31.3)), is much less than unity, and, consequently, ${\zeta}_{ef} \gg
{\zeta}_{a}$. An elongated ellipsoid resembles a cylinder. The case of an open
cylindrical Fermi surface is discussed in the next Subsection.  

\subsection{Open cylindrical Fermi surface}

The simplest model of an open Fermi surface is an infinitely long cylindrical
tube. Let the axis $z$ of the crystallographic coordinate system be the
cylindrical axis. The equation of such a surface in a cell of the momentum
space is 
$$
{\varepsilon}_{F} = \frac{{p}_{\bot}^{2}}{2m};\; -p_m < {p}_{z} < p_m,   
\eqno (35)
$$  
where $2p_m$ is the length of a single cell along the direction
$z$. In terms of the electron number density $n$ we have $p_m = 
n{\pi}^{2}{\hbar}^{3}/m{\varepsilon}_{F}$.  

For the cylindrical Fermi surface (35), the electron velocity ${\bf v}$ is in
the plane perpendicular to the axis $z$. Since $v_z = 0$, when calculating the
effective impedance we cannot use Eqs.(25) -(27). We have to repeat the
calculation beginning from the derivation of the proper expressions for the 
elements ${\sigma}_{\alpha\beta}(k;\gamma)$ of the conductivity tensor. 

First of all let's make use of Eqs.(17) and calculate the elements of the
conductivity tensor in the limit $kl \to \infty$. As before, we begin with the
calculation of the conductivity tensor with respect to the crystallographic
axes and use Eqs.(19.1) to calculate the tensor ${\sigma}_{ik}(k;\gamma)$ with
respect to the laboratory coordinate system. Again the result of the
calculation can be written as ${\sigma}_{ik}(k;\gamma) = {\sigma}_{a}(k)
{S}_{ik}$ with ${\sigma}_{a}$ defined by Eq.(14.2), where $S_F = 
4\pi p_m\sqrt{2m{\varepsilon}_{F}}$ is the lateral area of the
cylinder\footnote{The validity of the expression (14.1) for the averaged
conductivity in the case of an open cylindrical Fermi surface (35) can be
easily verified by averaging Eqs.(36) with respect to the Euler angles
${\varphi}_{k}$ and ${\theta}_{k}$.}. And the elements of the tensor ${S}_{ik}$
are 
$$
{S}_{11}=\frac{4}{\pi\sin {\theta}_{k}}\sin^2 {\varphi}_{k},\;
{S}_{12}=-\frac{2}{\pi\sin {\theta}_{k}}\sin 2{\varphi}_{k},\;
{S}_{22}=\frac{4}{\pi\sin {\theta}_{k}}\cos^2 {\varphi}_{k}. \eqno (36.1)
$$
and
$$
{S}_{13} = {S}_{23} = {S}_{33} = 0; \eqno (36.2)
$$
From Eqs.(36.1) it follows that in this approximation
${\sigma}_{11}{\sigma}_{22} - {\sigma}_{12}^{2} \equiv 0$. With regard to
Eqs.(22.2) this means that one of the principal values of the tensor 
${S}_{\alpha\beta}$, namely $S_2$, is equal zero. Next, if $S_2 = 0$, the
denominator in the expressions (21) for the Fourier coefficients of the
elements of the impedance tensor is $z(x=k{\delta}_{a};\gamma) = x^3
(x^3 - iS_1(\gamma))$ (see Eq.(22.1)), and the integrals (9) defining
the local impedance ${\zeta}_{\alpha\beta}$ diverge.

Thus we have shown, that for an open cylindrical Fermi surface in the limit 
$kl\to\infty$ the conductivity tensor has only one nonzero principal value. As
a result for such Fermi surfaces when calculating the surface impedance, we
cannot use the conductivity tensor (36). Consequently, we have to calculate the
tangential conductivities ${\sigma}_{\alpha\beta}(k;\gamma)$ up to the terms of
the order of $1/kl$. 
 
We use Eq.(13) and by analogy with Eq.(17.1) we write the elements of the
conductivity tensor in the form 
$$
{\sigma}_{ik} = {\sigma}_{a}(k){S}_{ik}(\gamma ;1/kl).
\eqno (37)
$$
The simple form of the Fermi surface allows us to calculate the elements of the
tensor ${S}_{ik}(\gamma ;1/kl)$ for an arbitrary value of the parameter $1/kl$.
With this result in hand it is easy to calculate the required principle values
${S}_{1}(\gamma ;1/kl)$ and ${S}_{2}(\gamma ;1/kl)$ up to the terms of the
order of $1/kl$. We omit the interim formulae and write down only the final
result:
$$
S_1 = \frac{4}{\pi\sin {\theta}_{k}}\left[1 - \frac{1}{kl\sin {\theta}_{k}}
\right]; \quad S_2 = \frac{4}{\pi kl}\cot^2 {\theta}_{k}.  \eqno (38)
$$

The anomalously small conductivity ${\sigma}_{2} = {\sigma}_{a}S_2$ contributes
to the leading terms in the expressions for the elements the local impedance
tensor ${\zeta}_{\alpha\beta}$. We use Eqs.(20) and Eqs.(21) to calculate the
Fourier coefficients of ${\zeta}_{\alpha\beta}(x;\gamma)$. It can be shown that
the additional small factor $1/kl$ in the expression for $S_2$ results in the
additional big factor ${(l/{\delta}_{a})}^{1/4}$ in the expressions for 
the elements of the local impedance tensor. Next, due to the same small factor,
the poles of the integrand of Eq.(9) are not the roots of the third-degree
equations (see Eq.(22.1)), but of the 4th-degree equation $x^4 - 4i\cot^2
{\theta}_{k}/\pi = 0$ (compare with the expression (38) for $S_2$). As a
result, for the cylindrical Fermi surface the relation between the real and the
imaginary parts of the surface impedance is not defined by the usual factor
$(1-i\sqrt{3})$. 

Taking into account the aforemention remarks, after calculating and consequent 
averaging of the elements of the local impedance tensor with respect to all
possible rotations of the crystallographic axes, we obtain the effective
impedance for the cylindrical Fermi surface:
$$
{\zeta}_{ef}^{(cyl)} = \frac{1}{8}\left(\frac{\omega{\delta}_{a}}{c} \right)
{\left(\frac{l}{4\pi {\delta}_{a}}\right)}^{1/4} {e}^{-3i\pi /8}
{\Gamma}^{2}(1/4),
\eqno (39). 
$$
where $\Gamma (x)$ is the gamma-function and ${\delta}_{a}$ is given by Eq.(16)
with $S_F = 4\pi p_m\sqrt{2m{\varepsilon}_{F}}$

We see that the absolute value of ${\zeta}_{ef}^{(cyl)}$ is much greater than
the typical value $|{\zeta}_{a}|$:
$$
\mid {\zeta}_{ef}^{(cyl)} \mid \sim \frac{\omega {\delta}_{a}}{c}
{\left(\frac{l}{{\delta}_{a}}\right)}^{1/4} \gg \mid {\zeta}_{a} \mid \sim
\frac{\omega {\delta}_{a}}{c}.   \eqno (40)
$$
We would like to point out that, as it was mentioned in the Introduction,
usually under the conditions of extremely anomalous skin effect, the elements
of the surface impedance tensor do not depend on the mean free path $l$. This
general conclusion is inapplicable for some specific Fermi surfaces. Our result
shows that it fails for an open cylindrical Fermi surface. The unusual factor 
${e}^{-3i\pi /8}$ in Eq.(39) is due to the aforemention unusual poles of the
integrand of Eq.(9). 

In the next Subsection, we show that in the case of a cubic Fermi surface (or
more generally, when the Fermi surface is a polyhedron ) the surface impedance
exhibits the same specific character.   

\subsection{Cubic Fermi surface}

Let the Fermi surface be a cube. Let the origin of the set of the
crystallographic axes be at the center of the cube. With respect to
crystallographic axes the sides of the cube are the planes
$$
{p}_{i}^{(a)} = \pm p_F \qquad (i = 1,2,3); \eqno (41)
$$
the edges of the cube are the intersections of the planes (41). At the sides of
the cube the velocity ${v}_{i}^{(a)} = \pm v_F$ (on the opposite sides the
directions of the vector $\bf v$ are opposite); the Fermi energy is
${\varepsilon}_{F} = {\bf vp}$.  

The surface (41) is not the surface of revolution and, consequently, Eqs.(27)
are not applicable. Moreover, it is evident, that for an arbitrary
direction of the wave vector $\bf k$ there are no "belts" on the cubic Fermi
surface where ${\bf kv} = 0$. This means that the approximation (17) for 
${\sigma}_{ik}(k;\gamma)$, as well as Eqs.(24), are not applicable either. So,
in this case, the starting point of our calculation is the general expression
(13) for the Fourier coefficients of the elements of the conductivity tensor.  

When the Fermi surface is a cube for an arbitrary value of the parameter 
$kl = kv_F\tau$, it is very easy to perform the integration in Eq.(13) with
respect to crystallographic coordinate system. It is evident that the only
nonzero elements of the tensor ${\sigma}_{ik}^{(a)}$ are its diagonal elements:
$$
{\sigma}_{qq}^{(a)} = \frac{4kl}{3\pi}{\sigma}_{a}\frac{1}{[1+
{\alpha}_{3q}^{2}{(kl)}^{2}]}, \quad q=1,2,3. \eqno (42.1)
$$
${\sigma}_{a}$ is given by Eq.(14.2) with $S_F$ being the lateral area of the
cube: $S_F=24p_F^2$. From Eq.(42.1) it is clearly seen that in this case, the
anisotropy of the conductivity tensor is due to the spatial dispersion only:
when $kl = 0$, the conductivity is a scalar. Next, the elements of the tensor
${\sigma}_{ik}(k;\gamma)$ with respect to the laboratory coordinate system are
$$
{\sigma}_{ik}(k; \gamma) = {\sigma}_{a}{S}_{ik}(kl; \gamma ), 
\qquad {S}_{ik}(kl; \gamma ) = \frac{4kl}{3\pi}\sum\limits_{q=1}^{3}
\frac{{\alpha}_{iq}{\alpha}_{kq}}{[1+{(kl)}^{2}{\alpha}_{3q}^{2}]}; 
\eqno (42.2)
$$
the elements of the rotation matrix ${\alpha}_{ik}$ are given by Eqs.(A1.1).

When $kl \gg 1$ from Eqs.(42) it follows, that for almost all the Euler angles
the first nonvanishing terms of the series expansion of all the elements of
${S}_{ik}$ in powers of the small parameter $1/kl$ are 
$$
{S}_{ik}(k;\gamma){|}_{kl\gg 1} = \frac{4}{3\pi}\frac{1}{kl}{\tilde F}_{ik};
\quad {\tilde F}_{ik} =\sum\limits_{q=1}^{3} 
\frac{{\alpha}_{iq}{\alpha}_{kq}}{{\alpha}_{3q}^{2}}.  \eqno (43)
$$
Thus, when $kl \gg 1$, for the cubic Fermi surface all the elements of the
conductivity tensor have the additional factor $1/kl$ and are much less than
the characteristic conductivity ${\sigma}_{a}$.  

It worth to be mentioned, that nevertheless the elements of the averaged
conductivity $<{\sigma}_{ik}({\bf k})>$ as before are given by Eqs.(14). The
point is that when averaging, Eqs.(43) lead to divergent integrals. In other
words, when averaging we cannot neglect $1$ in the denominator of the
expression (42) for ${S}_{ik}$. So, we have to use the general Eqs.(42) for an
arbitrary $kl$ to calculate the averaged conductivity and pass to the limit $kl
\to \infty$ after the integration is carried out.  

However, no divergence occurs when Eqs.(43) are used to calculate the surface
impedance (9). Then, with regard to Eqs.(10) - (12) we obtain the following 
expressions for the Fourier coefficients of the local surface impedance:
$$
{\zeta}_{\alpha\beta}(x;\gamma ) = -2i{\delta}_{a}
\left(\frac{\omega {\delta}_{a}}{c}\right)
\frac{x^2(x^4{\delta}_{\alpha\beta} - i\nu{f}_{\alpha\beta})}
{(x^4 - i\nu F_1)(x^4 - i\nu F_2)},  \eqno (44.1)
$$
where $x=k{\delta}_{a}$, ${\delta}_{a}$ is given by Eq.(16), $\nu =
4{\delta}_{a}/3\pi l \ll 1$ and 
$$
{f}_{11} = {\tilde F}_{22};\quad {f}_{12} = -{\tilde F}_{12}; \quad
{f}_{22} = {\tilde F}_{11};  \eqno (44.2)
$$
$$
F_{1,2} = \frac{1}{2}\left[{\tilde F}_{11} + {\tilde F}_{22} \pm
\sqrt{{({\tilde F}_{11} - {\tilde F}_{22})}^{2} + 4{\tilde F}_{12}^{2}}\right].
\eqno (44.3)
$$
(compare with the equations (21) and (22)). We present Eqs.(44) to show that
here, as in the case of an open cylindrical Fermi surface, the poles of the
Fourier coefficients ${\zeta}_{\alpha\beta}(x;\gamma )$ are the roots of 4th
degree equations $x^4 - i\nu {F}_{1(2)} = 0$. In addition, we would like to
mention, that the Fourier coefficients (44) depend on all of the three Euler
angles.  

When Eqs.(44) are substituted in Eq.(9) and the integration is carried out,
after the averaging with respect to the Euler angles we obtain the effective
impedance for the cubical Fermi surface: 
$$
{\zeta}_{ef}^{(cube)} = \frac{N}{4}\left(\frac{\omega {\delta}_{a}}{c}\right)
{\left(\frac{3\pi l}{{\delta}_{a}}\right)}^{1/4}{e}^{-3i\pi /8}, \quad 
N = <{F}_{1}^{-1/4} + {F}_{2}^{-1/4}>. \eqno (45)
$$
Numerical evaluation of the factor $N$ gives $N = 0.892$. We remind that
${\delta}_{a}$ is defined by Eq.(16) with $S_F = 24p_F^2$.

Of course, there are no cubic Fermi surfaces in real life. But there are metals
whose Fermi surfaces are close to polyhedrons (see, for example, ref.
\cite{20}). In this connection, let's estimate when smoothing of the edges and
the vertexes of the cube does not lead to a substantial change of the result
(45). Since the value of the local surface impedance (and, consequently, the
value of the effective impedance) is defined by the elements of the
conductivity tensor ${\sigma}_{ik}({\bf k})$, it is sufficient to estimate when
the contribution to ${\sigma}_{ik}({\bf k})$ from the smoothing regions is much
less than the contribution from the sides of the cube. We remind that the
contribution to the conductivity from the sides of the cube is given by
Eq.(42.2) and Eq.(43) with ${\sigma}_{a}$ from Eq.(14.2). When $kl \gg 1$ the
elements of the conductivity tensor (42.2) are of the order of 
${\sigma}^{(cube)}$,
$$
{\sigma}^{(cube)} \sim \frac{{e}^{2}{p}_{F}^{2}}{k(kl){\hbar}^{3}}.
$$

Let's estimate the contribution to the conductivity due to the smoothing of the
vertexes of the cube. Suppose $\delta {p}_{v}$ is the characteristic size of
the smoothing region, then with regard to Eqs.(17) the contribution to the
conductivity from the regions near the vertexes is of the order of 
$\delta{\sigma}^{(v)}$
$$
\delta{\sigma}^{(v)} \sim \frac{{e}^{2}{(\delta {p}_{v})}^{2}}{k{\hbar}^{3}}.
$$

Let's estimate the contribution to the conductivity due to the smoothing of the
edges of the cube. Here the characteristic size of the smoothing region in the
direction along an edge is of the order of ${p}_{F}$. Let $\delta {p}_{ed}$ be
the characteristic size of the smoothing region in the direction perpendicular
to the edge. Then with regard to Eqs.(17) the contribution to the
conductivity from the regions near the edges is of the order of 
$\delta{\sigma}^{(ed)}$
$$
\delta{\sigma}^{(ed)} \sim \frac{{e}^{2}{p}_{F}\delta {p}_{ed}}{k{\hbar}^{3}}.
$$

For a polycrystal with nearly cubic Fermi surface the effective impedance is
given by Eq.(45) when $\delta{\sigma}^{(v)}, \delta{\sigma}^{(ed)} \ll
{\sigma}^{(cube)}$. Then, if $\delta {p}_{ed} \sim \delta {p}_{v} \sim 
\delta p$, the value of $\delta p$ is limited by the inequality
$\delta{\sigma}^{(ed)} \ll {\sigma}^{(cube)}$, or
$$
\delta p \ll \frac{{p}_{F}}{kl}.   \eqno (46)
$$

Our result for the cubic Fermi surface, Eq.(45), is similar to Eq.(39) that is
the effective impedance in the case of the cylindrical Fermi surface. We have
shown that in both cases the value of the surface impedance is defined by small
conductivities of the order of ${\sigma}_{a}/kl$. Then the effective impedance
depends on the mean-free path $l$ and significantly exceeds the characteristic
value $|{\zeta}_{a}|$: $|{\zeta}_{ef}| \sim {(l/{\delta}_{a})}^{1/4}
|{\zeta}_{a}|$. In both cases the relation between the real and the imaginary
parts of the effective impedance is defined by unusual factor 
$\exp (-3i\pi /8)$.   
 
The unusual dependence of the surface impedance on the value of the mean-free
path $l$ obtained for the metals with cubic or cylindrical Fermi surfaces under
the conditions of extremely anomalous skin effect, is the result of direct
calculations. So, formally no further explanations are needed, but the
calculations are rather tedious and therefore the answer is not obvious. To
visualize the obtained result we use the Pippard method \cite{30}. A.B.Pippard
called it the method of ineffective electrons.  
 
Under the conditions of anomalous skin effect when calculating the surface
impedance the integral relation (7) between the current density $\bf j$ and
the electric field strength $\bf E$ has to be used instead of the Ohm low.  
According to Pippard, in the case of extremely anomalous skin effect the
correct result for the surface impedance can be obtained in the same way as
under the conditions of normal skin effect, if we take into account the fact
that the most part of the electrons at the Fermi surface is ineffective. Only a
small part of the electrons that is $\delta/l$ times less than in the case of
normal skin effect takes part in the reflection of electromagnetic waves.
A.B.Pippard used the standard local Ohm law, where the conductivity $\sigma 
\sim n{e}^{2}l/{p}_{F}$ was replaced by the effective value ${\sigma}_{P} \sim
\sigma (\delta/l)$. In this way the presence of "the belts" on the Fermi
surface was taken into account. Suppose, the introduction of the effective
conductivity ${\sigma}_{P}$ exhausts all the changes in the description of the
extremely anomalous skin effect compared with the case of normal skin effect.
Then we can calculate the penetration depth $\delta$ and the surface impedance
$\zeta$ with the aid of the standard equations substituting ${\sigma}_{P}$ for
$\sigma$. 

In other words, due to "the belts" on the Fermi surface, we cannot simply omit
$1$ in the denominator of the expression (13.1) for the conductivity 
${\sigma}_{ik}({\bf k})$. Now let's suppose that under the conditions of
extremely strong skin effect when calculating the principal values of the
transverse conductivity ${\sigma}_{\alpha\beta}({\bf k})$ for a given direction
of the wave vector $\bf k$, we can neglect $1$ without the divergence of the
integrals over the Fermi surface at least for one of the principal values.
Since the impedance is defined by the smaller of the principal conductivities,
the additional small factor $\delta/l$ appears in the effective conductivity
${\sigma}_{P}$: ${\sigma}_{P} \sim \sigma {(\delta/l)}^{2}$. When the Pippard
method is used, namely this effective conductivity provides the correct value
of the impedance. Now it is evident that 
$$
\delta \sim \frac{c}{\sqrt{i{\sigma}_{P}\omega}} \sim 
{\left(\frac{{c}^{2}{l}^{2}}{i\sigma\omega}\right)}^{1/4}; \qquad \zeta \sim 
\frac{\omega\delta}{c}.
$$
Then, with regard to the last equation $\zeta \sim l^{1/4}$. Note that the
Pippard method allows us to define all the dimensional factors correctly, as
well as the relation between real and imaginary parts of the impedance.

If for a single crystal metal this situation takes place for a finite interval
of the directions of the wave vector $\bf k$ (or, in other words, for a finite
interval of the Euler angles $\gamma$), the relation $\zeta \sim l^{1/4}$
remains valid for the polycrystal too. The last is true since when averaging
the leading term is defined by the Euler angles corresponding to the maximal
values of the local impedance. Since all the conditions mentioned above are
realized for polycrystals composed of the single crystal grains with cubic or
cylindrical Fermi surfaces, their impedance has to be proportional to 
${l}^{1/4}$. The last conclusion is consistent with Eq.(39) and Eq.(45).
\vskip 3mm 
\section{EFFECTIVE IMPEDANCE IN VICINITY OF ELECTRONIC TOPOLOGICAL
TRANSITION} 

The possibility to observe the effect of a change of the topology of the Fermi
surface on the properties of electrons was predicted by I.M.Lifshitz \cite{21}.
The change of the topology of the Fermi surface takes place when the Fermi
energy equals to one of the critical values ${\varepsilon}_{c}$ determined by
band edges, local maxima and minima of the function $\varepsilon ({\bf p})$ and
the Van Hove singularities. If the Fermi surface is found in the vicinity of a
critical point, it can be slightly "corrected" by some external effect (e.g.
applying pressure or adding some impurities). The change in the Fermi surface
topology is called the electronic topological transition. As a consequence of
such a change, the properties of a metal determined by the Fermi surface
electrons, exhibit the singularities with different critical exponents. In
particular, the singular addition to the thermodynamic potential at zero
temperature is of the order of ${|{\varepsilon}_{F}-{\varepsilon}_{c}|}^{5/2}$.
This allowed I.M.Lifshitz to call the electronic topological transition as the
$2\frac{1}{2}$ order phase transition in accordance with Ehrenfest's
terminology.    

Two basic types of topological transitions are possible depending on the type
of the critical point that the Fermi surface passes through. They are: 

\noindent 1. the formation of a new void of the Fermi surface or the
disappearance of an existing void when the critical point corresponds to an
extremum of the function ${\varepsilon}({\bf p})$; 

\noindent 2. the creation or the disruption of a neck when the critical point
corresponds to a conic point of the Fermi surface. 

\noindent Some exotic cases are also possible. For example, the critical points
form a curve in the momentum space. This situation takes place for wurzite type
crystals (see Subsection V.B below). An attempt to review theoretical papers
devoted to the electronic topological transition has been done in ref.
\cite{22}. Here the detailed bibliography on the subject can also be found. 

Kinetic properties of metals are especially sensitive to the Fermi surface
structure. For example, the conductivity exhibits anomalies near the
topological transition. The analysis shows (see ref. \cite{22}) that if the 
spatial dispersion of the conductivity is of no importance, the main influence
of the topological transition on the value of the conductivity is indirect: the
anomaly is related to the variation in the electron mean free path $l$.
This is not the case under the conditions of extremely anomalous skin
effect, when the electron scattering is less important than the specifics of
the motion of free electrons with complicated dispersion relation. In this
case, the sensitivity of the kinetic properties to the structure of the Fermi
surface defines their strong dependence on the parameter ${\varepsilon}_{F}-
{\varepsilon}_{c}$.   

There are papers, where under the conditions of anomalous skin effect the
surface impedance has been calculated for single crystal metals with complex
Fermi surfaces (see, for example, refs. \cite{23,24}). Also there are
investigations of the surface impedance of single crystals in the vicinity
of the electronic topological transition. The question we are investigating in
this Section is: whether the singularities related to the electronic
topological transition "survive" in polycrystals, which, in effect, are
isotropic metals. In what follows we show that the singularities do "survive":
the effective surface impedance of the polycrystal exhibits nontrivial behavior
in the vicinity of the electronic topological transition. 

Even without the calculation it is easy to understand that when a new little
void of the Fermi surface appears, the derivative of the effective impedance
has a jump. Really, suppose we examine a single crystal metal in the vicinity
of a critical point corresponding to an extremum of the function 
${\varepsilon}({\bf p})$. Then the newly appearde void is an ellipsoid. When
calculating the singularity of the surface impedance under the conditions of
extremely strong skin effect, it is usual to assume that even for electrons of
the small void the mean free path $l \to \infty$. Next, the surface impedance
is proportional to ${S}_{F}^{-1/3}$; ${S}_{F}$ is the total area of the Fermi
surface. After the formation of the new void ${S}_{F} = {S}_{0} + {S}_{v}$,
where ${S}_{0}$ is the area of the main part of the Fermi surface and ${S}_{v}$
is the area of the new ellipsoidal void: ${S}_{v} \sim |{\varepsilon}_{F} -
{\varepsilon}_{c}|$. Consequently, in the vicinity of the critical point the
impedance $\zeta = {\zeta}_{0} +  \delta\zeta$, where $\delta\zeta = 0$ until
the formation of the void, and $\delta\zeta \sim {S}_{v}$ is the addition to
the impedance caused by the appearance of the new void. Thus, in the case of
single crystal metal, the derivative of the impedance has a jump when
${\varepsilon}_{F} = {\varepsilon}_{c}$. 

When calculating the effective impedance of the polycrystal, we have to average
the surface impedance of a single crystal with respect to all possible
rotations of the crystallographic axes with respect to the fixed set of
laboratory axes. Since in the vicinity of the topological transition the
singular addition to the surface impedance caused by the formation of a new
ellipsoidal void has the same sign and is of the same order for all the
directions of the crystallographic axes, the averaging does not change the
result and the derivative of ${\zeta}_{ef}$ also has a jump when
${\varepsilon}_{F} = {\varepsilon}_{c}$.

If the topological transition leads to the creation or the disruption of a neck
of the Fermi surface, the character of the effective impedance singularity
cannot be obtained without the calculation. The point is that the orientation
of the neck defines "a preferred direction", and the averaging strongly affects
the value of the surface impedance. In what follows, we examined a polycrystal
composed of the single crystal grains with the Fermi surface of a goffered
cylinder type. This example provides a rather general description of the
singularity near the conic point. Usually under the topological transition, the
singularities of thermodynamic and kinetic characteristics depend only on the
structure of the Fermi surface in the vicinity of the point where the change of
the topology occurs. Our calculation does not contradict this rule. As we
showed, the singular addition to the effective impedance depends only on the
ratio of the effective masses at the point of the topological transition. 

The other example we examined was the polycrystal composed of the single
crystal grains with the Fermi surface of a wurzite crystal type; the Fermi
energy is close to the critical value corresponding to the disappearance of the
toroidal hole of the Fermi surface and the appearance of the new ovaloid void. 
This example is an exotic one. It has been chosen following ref.\cite{22}
because of the relative simplicity of analytical analysis.

In conclusion of this Subsection one remark must be done. If the external
pressure is used to observe the effect, it must be taken into account that in
a polycrystalline sample, the stresses can be different in different grains.
Then the transition would be blurred. It is better to use polycrystals where
the inhomogeneity of the stresses is minimal. May be the results of ref.
\cite{32} will be useful for the choice of such polycrystals.

\subsection{Fermi surface of goffered cylinder type}

Let the polycrystal be composed of single crystal grains whose Fermi surface
with regard to the crystallographic axes is defined by the equation 
$$
{\varepsilon}_{F} = \frac{{p}^{2}_{\bot}}{2{m}_{\bot}} + 
{\varepsilon}_{c}\cos\frac{\pi{p}_{z}}{p_m},  \eqno (47)
$$
The length of the unit cell in the direction $z$ is $2p_m$. For the
crystallographic axes we choose the origin of coordinates at the point $p_z =
0$, then $-p_m < p_z < p_m$. The topology of the Fermi surface changes at the
point ${p}_{z}=0$ when ${\varepsilon}_{F}$ equals to the critical value
${\varepsilon}_{c}$. Namely, 

\noindent 
1) if ${\varepsilon}_{F} < {\varepsilon}_{c}$, the Fermi surface is a
closed surface (under our choice of the unit cell, in each cell there are two
separated parts belonging to two different closed surfaces, Fig.2.a);  

\noindent 
2) if ${\varepsilon}_{F} = {\varepsilon}_{c}$, the Fermi surface has
a conic point and a neck is formed (see Fig.2.b);

\noindent 
3) if ${\varepsilon}_{F} > {\varepsilon}_{c}$, the Fermi surface is
an open one (see Fig.2.c).

Having in mind to calculate the change of the effective surface impedance in
the vicinity of the electronic topological transition, we set 
$$
{\varepsilon}_{F}/{\varepsilon}_{c} = 1 + \delta\varepsilon.   \eqno (48)
$$
The cases $\delta\varepsilon < 0$ and $\delta\varepsilon > 0$ must be examined
separately. 

Since the area of the Fermi surface $S_F$ varies under the variation of the
Fermi energy ${\varepsilon}_{F}$, it is not convenient to use ${\zeta}_{a}$
(see Eq.(16)) as the characteristic value of the surface impedance. Let us
introduce new characteristics of the Fermi surface:
$$
p_c = \frac{{m}_{\bot}{\varepsilon}{c}}{p_m}; \quad S_c = 4\pi{p}_{c}^{2}; 
\quad \gamma = {\pi}^{2}\frac{p_c}{p_m}, \eqno (49)
$$
The values of ${p}_{c}, {S}_{c}$ and $\gamma$ do not depend on the Fermi
energy. Now we rewrite Eqs.(27) for the effective impedance in the form: 
$$
{\zeta}_{ef} = {\zeta}_{g}Z(\delta\varepsilon ;\gamma). \eqno (50.1)
$$
Here, by analogy with Eq.(16) we introduce
$$
{\zeta}_{g} = \frac{2(1-i\sqrt{3})}{3\sqrt{3}}
\left(\frac{\omega{\delta}_{g}}{c}\right),\quad
{\delta}_{g} = {\left(\frac{4\pi{c}^{2}{\hbar}^{3}}
{\omega{e}^{2}{S}_{c}}\right)}^{1/3}.
\eqno (50.2)
$$
It is evident that ${\zeta}_{g}$ does not depend on ${\varepsilon}_{F}$ either.
In other words, in the vicinity of the conic point the value of ${\zeta}_{g}$
is fixed. Next, if instead of the 
Euler angle ${\theta}_{k}$ we introduce $w =  \tan^2 {\theta}_{k}/\gamma$, from
Eq.(27.2) it follows that the function $Z(\delta\varepsilon ;\gamma)$ is
$$
Z(\delta\varepsilon ;\gamma) = \frac{\gamma}{2{4\pi}^{1/3}}
\int\limits_0^{\infty}\frac{[{\tilde S}_{1}^{-1/3} + {\tilde S}_{2}^{-1/3}]}
{{(1+\gamma w)}^{3/2}} dw. \eqno (50.3)
$$
In Eq.(50.3) the functions ${\tilde S}_{1(2)}$ depend on $w, \gamma$ and
$\delta\varepsilon$. They are the renormolized functions $S_1$ and $S_2$
defined by Eq.(25.1) and Eq.(25.2) respectively. If, when integrating over
${p}_{\bot}$ in Eqs.(25) the dimensionless variable $x ={p}_{\bot}/\pi{p}_{c}$
is used, we obtain  
$$
{\tilde S}_{1}(w;\gamma,\delta\varepsilon) = \frac{\sqrt{1+w\gamma}}
{w\gamma}\int x \Phi (x;w,\delta\varepsilon)dx; \quad
{\tilde S}_{1}(w;\gamma,\delta\varepsilon) = \frac{{(1+w\gamma)}^{3/2}}
{w\gamma}\int \frac{xdx}{\Phi (x;w,\delta\varepsilon)},
$$
where
$$
\Phi (x;w,\delta\varepsilon)) = \sqrt{\frac{\gamma w{x}^{2 }}
{1-{(1+\delta\varepsilon - \gamma{x}^{2}/2)}^{2}} - 1}.    \eqno (50.4)    
$$
The requirement for the radicand of Eq.(50.4) to be positive combined with
inequalities defining the intervals of $x$ variation on the Fermi surface give
the region of integration.  

Let ${S}_{1(2)}^{(c)}$ be the functions ${\tilde S}_{1(2)}$ for
$\delta\varepsilon = 0$ (the Fermi energy corresponds to the conic point). Let
${\zeta}_{ef}^{(c)}$ be the effective impedance relevant to this Fermi energy.
According to Eqs.(50) we have
$$
{\zeta}_{ef}^{(c)} = {\zeta}_{g}{Z}_{c}(\gamma);\quad {Z}_{c}(\gamma) =
Z(\delta\varepsilon = 0;\gamma).  \eqno (51)
$$
The function ${Z}_{c}(\gamma)$ is presented in Fig.3.

Let $\delta{\zeta}_{ef}$ be the difference between the value of the effective
impedance when $\delta\varepsilon \ne 0$ and ${\zeta}_{ef}^{(c)}$:
$\delta{\zeta}_{ef}(\delta\varepsilon) = {\zeta}_{ef}(\delta\varepsilon) - 
{\zeta}_{ef}^{(c)}$. It is evident that if $|\delta\varepsilon | \ll 1$ in the
main approximation
$$
\delta{\zeta}_{ef}(\delta\varepsilon) = {\zeta}_{g}[\delta z_1
+ \delta z_2];   \eqno (52.1)
$$
$$
\delta z_{\alpha}(\delta\varepsilon ) \approx \frac{\gamma}{6{(4\pi)}^{1/3}}
\int\limits_{0}^{\infty}\frac{dw}{{(1 + \gamma w)}^{3/2}}
\frac{\Delta {\tilde S}_{\alpha}}{{[{S}_{\alpha}^{(c)}]}^{4/3}}, \quad 
\alpha \;=\;1,2; \eqno (52.2)
$$
and
$$
\Delta {\tilde S}_{\alpha} = {S}_{\alpha}^{(c)}(w;\gamma) - 
{\tilde S}_{\alpha}(w;\gamma,\delta\varepsilon).  \eqno (52.3)
$$

Below we present the results obtained when calculating 
$\delta{\zeta}_{ef}(\delta\varepsilon)$ for $\delta\varepsilon < 0$ and 
$\delta\varepsilon > 0$. The details of the calculation are straightforward,
but rather lengthy and tedious. In Appendix 3 we outline the main steps of the
calculation and write down the main interim results. The expressions for
$\delta{\zeta}_{ef}(\delta\varepsilon)$ are different depending on the sign of
$\delta\varepsilon$. In what follows we use the superscript $(<)$ to
mark all the values calculated for $\delta\varepsilon < 0$ and the superscript
$(>)$ to show that $\delta\varepsilon > 0$. 

So, when $|\delta\varepsilon| \ll 1$ our final
result for $\delta{\zeta}_{ef}^{(<)}$ is
$$
\delta{\zeta}_{ef}^{(<)} \approx \frac{{\zeta}_{g}}{3}
{\left(\frac{\gamma}{2\pi}\right)}^{5/3}\frac{{\alpha}^{(<)}}
{{(1+\gamma )}^{2}}{(|\delta\varepsilon |/2)}^{3/4}\ln |\delta\varepsilon |,
\eqno (53.1)
$$
where
$$
{\alpha}^{(<)} = \int\limits_0^1\frac{(1+x^4)dx}{\sqrt{1-x^4}}.  \eqno (53.2)
$$
Numerical evaluation gives ${\alpha}^{(<)} \approx 1.75$. 

When $\delta\varepsilon > 0$ our final result for $\delta{\zeta}_{ef}^{(>)}$ is
$$
\delta{\zeta}_{ef}^{(>)} \approx -\frac{{\zeta}_{g}\gamma}{3}
{\left(\frac{\gamma}{2\pi}\right)}^{5/3}
\frac{{\alpha}^{(>)}}{{(1+\gamma )}^{2}}
{(\delta\varepsilon /2)}^{3/4}\ln \delta\varepsilon,  \eqno (54.1)
$$
where
$$
{\alpha}^{(>)} = \int\limits_0^1
\frac{(1-x^4)dx}{\sqrt{1+x^4}}.   \eqno (54.2)
$$
Numerical evaluation gives ${\alpha}^{(>)} \approx .77$. 

The equations (53) and (54) describe the singularity of the effective impedance
of polycrystals in the vicinity of the electronic topological transition when
the Fermi surface passes through the conic point (Fig.4). Firstly, we would
like to note, that the dependence of ${\zeta}_{ef}$ on the parameter 
$\delta\varepsilon = {\varepsilon}_{F}/{\varepsilon}_{c} -1$ is nontrivial: 
${\delta\zeta}_{ef} \sim {|\delta\varepsilon |}^{3/4}\ln |\delta\varepsilon|$. 
Thus, in the case under consideration, the sigularity of the effective
impedance is stronger than when a new void appears: at the conic point the
derivative $\partial {\zeta}_{ef}/\partial {\varepsilon}_{F}$ has the
infinitely large jump. 

In conclusion, let us show that the character of the singularity, the value of
the ratio $\delta {\zeta}_{ef}/{\zeta}_{g}$, depends on the structure of the
Fermi surface near the conic point only. Indeed, let's introduce the effective
masses ${m}_{\bot}^{ef}={\partial}^{2}\varepsilon ({\bf p})/\partial 
{p}_{\bot}^{2}$ and ${m}_{z}^{ef}={\partial}^{2}\varepsilon ({\bf p})/\partial
{p}_{z}^{2}$. It is evident that everywhere at the Fermi surface 
${m}_{\bot}^{ef} = {m}_{\bot}$. In our case, the change of the topology takes
place in the point ${\bf p} = 0$ of the momentum space. Here 
${m}_{z}^{ef} = {p}_{m}^{2}/{\pi}^{2}{\varepsilon}_{c}$. In the 
small vicinity of the point ${\bf p} = 0$ we can expand the expression (47) and
write the equation of the Fermi surface as 
$$
{\varepsilon}_{F} -{\varepsilon}_{c} = \frac{{p}^{2}_{\bot}}{2{m}_{\bot}} -
\frac{{p}^{2}_{z}}{2{m}_{z}^{ef}}.
$$
Next, with regard to Eq.(49) the parameter $\gamma$ in the expressions (53) and
(54) for $\delta {\zeta}_{ef}/{\zeta}_{g}$ is simply the ratio of the effective
masses: $\gamma = {m}_{\bot}/{m}_{z}^{ef}$ at the point ${\bf p}=0$.
Consequently, only the local geometry of the Fermi surface in the vicinity of
the conic point affects the singularity of the effective impedance.  

\subsection{Fermi surface of wurzite type crystals} 

Let the polycrystal be composed of single crystal grains with an energy
spectrum of wurzite type crystals \cite{17}. The dispersion relation for such
crystals written down with respect to the crystallographic coordinate system is
$$
{\varepsilon}^{\pm}({\bf p}) = \frac{1}{2{m}_{\bot}}
{({p}_{\bot} \pm {p}_{0})}^{2} + \frac{1}{2m_z}{p}^{2}_{z},   \eqno (55)
$$
where the effective masses ${m}_{\bot}$ and $m_z$ are positive. 

There are two critical energies ${\varepsilon}_{c}^{\pm}$ corresponding to the
change of topology of the equienergy surface (55). The first critical value of
the energy ${\varepsilon}_{c}^{-}$ (we set ${\varepsilon}_{c}^{-} = 0$)
corresponds to the appearance of a new void of the equienergy surface related
to minus sign in  Eq.(55). When ${\varepsilon}_{F} = {\varepsilon}_{c}^{-}$,
the critical points ${p}_{\bot} = p_0$ form the curve in the momentum space. If
$0 < \varepsilon < {\varepsilon}_{c}^{+}$, where ${\varepsilon}_{c}^{+} =
p^2_0/2{m}_{\bot}$, the equienergy surfaces are toroids with elliptical
crossection in the planes containing the axis $p_z$ (Fig.5.a). When the energy
equals to the second critical value ${\varepsilon}_{c}^{+}$, the hole in the
toroid disappears and a new void of ovaloid shape related to plus sign in
Eq.(64) appears (Fig.5.b) at the point ${p}_{\bot} = p_z = 0$. Thus, the 
point ${p}_{\bot} = p_z = 0$ is an isolated critical point. 

In what follows for $0< {\varepsilon}_{F} < {\varepsilon}_{c}^{+}$ we calculate
${\zeta}_{ef}={\zeta}_{ef}^{(<)}$ as a function of
${\varepsilon}_{F}/{\varepsilon}_{c}^{+}$. For ${\varepsilon}_{F} >
{\varepsilon}_{c}^{+}$ we write down the value of the effective impedance
${\zeta}_{ef}^{(>)}$ as a sum:
$$
{\zeta}_{ef}^{(>)}({\varepsilon}_{F}/{\varepsilon}_{c}^{+}) =  
{\zeta}_{ef}^{(<)}({\varepsilon}_{F}/{\varepsilon}_{c}^{+}) + \Delta\zeta.
\eqno (56)
$$
Here ${\zeta}_{ef}^{(<)}({\varepsilon}_{F}/{\varepsilon}_{c}^{+})$ is the
extension of the function ${\zeta}_{ef}^{(<)}$ to the region 
${\varepsilon}_{F}/{\varepsilon}_{c}^{+} >1$, and $\Delta\zeta$ describes the
singularity of the effective impedance in the vicinity of the topological
transition related to the critical energy ${\varepsilon}_{c}^{+}$. 

We performed the calculation with the aid of Eqs.(25) - (27). Some main interim
formulae are presented in Appendix 4. As in the previous Subsection, here it is
reasonable to introduce a characteristic impedance ${\zeta}_{w}$ which does not
change with the change of the Fermi energy. With regard to the equation of the
Fermi surface (55) we set 
$$
{\zeta}_{w} = \frac{2(1-i\sqrt{3})}{3\sqrt{3}}
\left(\frac{\omega{\delta}_{w}}{c}\right),\quad
{\delta}_{w} = {\left(\frac{\pi{c}^{2}{\hbar}^{3}}
{\omega{e}^{2}{m}_{\bot}{\varepsilon}_{c}^{+}}\right)}^{1/3}.  \eqno (57)
$$

\underline {Let's begin with the case $0< {\varepsilon}_{F} < 
{\varepsilon}_{c}^{+}$}. Then the Fermi surface is the toroid 
${\varepsilon}_{F} = {\varepsilon}^{-}({\bf p})$ (see Eq.(55)). The direct
calculations show that for such values of the Fermi energy the effective
impedance is 
$$
{\zeta}_{ef}^{(<)}({\varepsilon}_{F}/{\varepsilon}_{c}^{+}) =
{\zeta}_{w}{\left(\frac{{\varepsilon}_{c}^{+}}{{\varepsilon}_{F}}\right)}^{1/6}
B\left(\frac{m_z}{{m}_{\bot}}\right),   \eqno (58.1)
$$
where the function $B(z)$ is defined by Eqs.(A4.3) of Appendix 4. For $z \ll 1$
and $z \gg 1$ we have
$$
B(z) \approx \frac{5}{16}{\left(\frac{4}{\pi z}\right)}^{1/3}, \;{\rm if}\; 
z \ll 1
$$
$$
B(z) \approx \frac{\pi}{8}\frac{{z}^{1/6}}{\ln z}, \;{\rm if}\; z \gg 1. 
\eqno (58.2)
$$

We see that the effective impedance increases if the Fermi energy
${\varepsilon}_{F}$ decreases. According to Eq.(58.1) ${\zeta}_{ef}^{(<)}
\gg {\zeta}_{w}$, when ${\varepsilon}_{F} \to 0$. We remind that we have chosen
the first critical value of the energy, ${\varepsilon}_{c}^{-}$, to be equal
to zero. When ${\varepsilon}_{F} = 0$, the Fermi surface is not a three
dimensional surface, but the ring ${p}_{\bot} = {p}_{0}; p_z = 0$. Evidently,
on such a surface there are no "belts", where the condition ${\bf kv} = 0$ is
fulfilled. Consequently, the conductivity is unusualy small and the impedance
is extremely large. Next, the effective impedance increases significantly in
the cases of strongly flattened (${m}_{z}/{m}_{\bot} \ll 1$) or strongly
elongated (${m}_{z}/{m}_{\bot} \gg 1$) toroids. The situation is very similar
the one discussed in Subsection IV.A, where we have shown that the effective
impedance is extremely large in the cases of strongly flattened and strongly
elongated ellipsoids. Generally it can be stated that the surface impedance
increases significantly, when "the dimension" of the Fermi surface decreases.  

\underline {When ${\varepsilon}_{F} > {\varepsilon}_{c}^{+}$}, the Fermi
surface consists of two parts: the external one is defined by the equation
${\varepsilon}_{F} = {\varepsilon}^{-}({\bf p})$, and it is a part of the
toroid; the internal one is given by the equation ${\varepsilon}_{F} = 
{\varepsilon}^{+}({\bf p})$, and it is a part of the ovaloid (see Eq.(55)). 

With regard to Eq.(56) the straightforward calculation of the function
$\Delta\zeta$ shows that near the point of the topological transition, 
i.e. when $0 < \delta\varepsilon = ({\varepsilon}_{F} -
{\varepsilon}_{c}^{+})/{\varepsilon}_{c}^{+} \ll 1$, we have 
$$
\Delta\zeta = -{\zeta}_{w}{\delta\varepsilon}^{3/2}
C\left(\frac{m_3}{m_\bot}\right),  \eqno (59.1)
$$
where the function $C(z)$ is defined by Eqs.(A4.5) of Appendix 4. For $z \ll 1$
and $z \gg 1$ we have 
$$
C(z) \approx \frac{1}{36\sqrt{\pi}{z}^{5/6}}
{\left(\frac{1}{2\pi}\right)}^{1/3}\frac{\Gamma (1/6)}{\Gamma (2/3)}, 
\qquad {\rm if}\; z \ll 1;
$$
$$
C(z) \approx \frac{1}{120{z}^{1/6}}, \qquad {\rm if}\; z \gg 1;
\eqno (59.2)
$$
$\Gamma (x)$ is the gamma-function. 

Equations (59) describe the singularity of the effective surface impedance
related to the disappearance of the hole of the toroidal Fermi surface and the
appearance of a new ovaloid void. We would like to note that the singularity
related to the new void formation at the isolated critical point
${\varepsilon}_{c}^{+}$ (see Eq.(59.1)), is weaker than in the case of a neck
formation (see Eqs. (53.1), (54.1)). It is also weaker than in the case of the
new void appearance in the vicinity of an extremum of the function 
$\varepsilon ({\bf p})$.
\vskip 3mm

\section{CONCLUSIONS}

The calculation of the surface impedance of polycrystalline metals is a logical
result of the development of electron theory of metals. The possibility to
calculate the surface resistance of a polycrystalline metal with high accuracy
(in fact, exactly) when the impedance (the Lentovich) boundary conditions are
valid \cite{12}, was a stimulus to investigate all the kinds of different
physically meaningful situations. Anomalous skin effect is on of these
situations. The present investigation together with refs. \cite{6,7,8} and
refs. \cite{10,11} completes the theory of skin effect in polycrystals. 

Polycrystals are inhomogeneous solids, and all inhomogeneous media greatly
interested scientists in the past years. A distinctive feature of a polycrystal
is its grains. The grains, being the structural elements of polycrystals are
macroscopic formations. The inhomogeneity of polycrystals is due to random
orientation of the crystallographic axes of the grains with respect to each
other. As a result, when electron properties of single crystal grains out of
which the polycrystal is composed are known, not is it only possible to set up
a problem of the effective impedance calculation, but also to solve it too. The
region of the obtained results application is very wide. They are applicable
when the surface resistance is of interest. For example, if the reflectance and
absorbtion of metallic mirrors, which are elements of modern devices and
systems, are under investigation. 

Suppose the impedance of an individual single crystal grain is its
phenomenological characteristic corresponding to a flat metal-vacuum interface,
then Eq.(5) has to be considered as the final formula of the theory. For
metallic polycrystals this formula has been obtained in ref. \cite{10}; with
regard to the radio waves propagation along the earth surface, a similar result
has been obtained by E.L.Feinberg as far back as in the mid forties. Comparison
of Eq.(5) with other formulae obtained for effective characteristics with the
aid of the method of I.M.Lifshitz and L.N.Rosenzweig \cite{1}, shows the
principal difference from the formula for the effective impedance obtained in
the framework of the Leontovich boundary conditions applicability. The point is
that the spatial correlators of the random characteristics of the polycrystal
do not enter Eq.(5). Taking account of these correlators falls outside the
limits of the accuracy of the Leontovich boundary conditions. This allows us to
consider Eq.(5) as an exact formula.

To start the calculation with the aid of Eq.(5), it is necessary to know the
impedance of the single crystal metal for an arbitrary orientation of the
crystallographic axes with respect to the metal surface. Without simplifying
assumptions the impedance of single crystals can be calculated in two limiting
cases only. Namely, under the conditions of normal skin effect (the impedance
is expressed in terms of the elements of the conductivity tensor \cite{10,11})
and under the conditions of extremely strong skin effect. In the last case the
impedance is expressed in terms of integrals over the Fermi surface, and
usually it does not depend on dissipative characteristics of the metal
\footnote{We would like to note, that in the intermediate case, when $\delta$
is of the order of $l$, an analytic expression for the impedance can be
obtained only after rather significant simplifications (see, for example, 
\cite{18}).}.

Under the conditions of extremely strong skin effect Eqs.(24) together with
Eq.(17.2) solve the problem of the effective impedance calculation. Later on
(except for Section V) Eqs.(24) are analyzed for some model or "exotic" Fermi
surfaces. If for the given Fermi surface all the principle values of the tensor
(17.2) are of the order of unity, the values of the impedance of the single
crystal, calculated for an arbitrary orientation of the sample surface, are of
the same order. The effective impedance is of the same order too. In this
case the obtained results do not change the picture of the anomalous skin
effect qualitatively. But we must take into account that the averaging (the
angular brackets in Eq.(24.2)) changes the value of the impedance
quantitatively.  

Often the results of the measurement of the impedance under the conditions of
extremely strong skin effect are used to calculate the electron mean free path
in polycrystals. The calculation is carried out in the following way. Suppose
the polycrystal is an isotropic conductor with a spherical Fermi surface
${S}_{F}^{(a)}$, then its impedance is defined by Eq.(16). The measurement of
the impedance allows us to calculate ${S}_{F}^{(a)}$. Next, the static
conductivity of an isotropic conductor is $\sigma \sim {S}_{F}^{(s)}l$, where
${S}_{F}^{(s)}$ is again the area of the Fermi surface. Suppose ${S}_{F}^{(a)}
\approx {S}_{F}^{(s)}$, then the value of the mean free path $l$ can be
calculated after the measurement of the specific resistance $\rho = 1/\sigma$.

Of course, this method can be used, if the anisotropy of the single crystal
grains is small. However, if the anisotropy is strong, the results of the
measurements must be handled with care. First of all, our results shows that
the area ${S}_{F}^{(a)}$ is not the real area of the Fermi surface: the
averaging in Eqs.(24) gives a numerical factor that can significantly differ
from unity. Then the measured value of ${S}_{F}^{(a)}$ is {\it the effective
area of the Fermi surface related to anomalous skin effect}. Next, the
effective static conductivity of a strongly anisotropic polycrystal does not
equal to the static conductivity, averaged over all possible rotations of the
grains. The difference can be very big (see ref. \cite{33}). Then 
${S}_{F}^{(s)}$ that enters the equation for the static condudtivity is an
effective area defined with regard to the static conductivity. Of course,
${S}_{F}^{(s)} \ne {S}_{F}^{(a)}$ (see ref. \cite{11}, where the effective
static conductivity of two-dimensional polycrystals has been compared with the
effective conductivity related to normal skin effect), and this difference has
to be taken into account when estimating the electron mean free path in
polycrystals.  

When the Fermi surface is an axially symmetric surface, the expression for the
effective impedance is much simpler, since the explicit form of the principal
values ${S}_{1}$ and ${S}_{2}$ of the tensor ${S}_{ik}$ (see Eq.(17)) can be
obtained (see Eqs. (25-27)).   

We would like to point out that the Fermi surfaces of the majority of real
metals are extremely complex. They have many voids of different shapes and
symmetry. If one of the voids is axially symmetric, but not all the others,
Eqs.(25)-(27) are inapplicable. We can use Eqs.(25-27) only if the Fermi
surface is axially symmetric as a whole.

When calculating the effective impedance for some different model Fermi
surfaces (Section IV), several problems were the object of our investigation.
First of all, the analysis of an ellipsoidal Fermi surface is the first
necessary step, when after the analysis of an isotropic conductor (the
conductor with the spherical Fermi surface), we turn to the investigation of
real anisotropic polycrystals, but the results of Subsection IV.A also can be
applied to bismuth type semimetals and some degenerate semiconductors. In this
case, however, these results have to be generalized, since as a rule the Fermi 
surfaces of semimetals and degenerate semiconductors are the sets of
ellipsoids. The discussed cases of extremely anisotropic ellipsoids, besides
being an illustration of the change of the effective area of the Fermi surface
(see Eq.(33.2) and Eq.(33.3)), can be helpful when low-dimensional systems are
analyzed: Eq.(31.2) and the following ones where $\mu \ll 1$ are relevant for
the description of quasionedimensional conductors; Eq.(31.3) and the following
ones where $\mu \gg 1$ are relevant for the description of quasitwodimensional
conductors.  

The analysis of metals with cylindrical (Subsection IV.B) and cubic (Subsection
IV.C) Fermi surfaces shows that the effective impedance of polycrystals can
differ from the impedance of an isotropic metal {\it qualitatively}. By a
qualitative difference we mean the dependence of ${\zeta}_{ef}$ on the mean
free path $l$ under the conditions of extremely anomalous skin effect (see
Eq.(39) and Eq.(45)). On the other hand, the results of Section IV can be used
as the first approximation under the description of real polycrystals. In
particular, the Fermi surfaces of quasitwodimensional metals are slightly
goffered cylinders. Eq.(39) can be applicable to such metals if the binding of
the conducting planes is weak and the goffering can be neglected. Or, if the
goffering can be considered as a perturbation, Eq.(39) is the zeroth term of
the perturbation theory. 

We do not know whether there are metals whose Fermi surfaces are close to a
cube as a whole. Apparently, there are metals with a nearly cubic main void of
the Fermi surface (see ref. \cite{20}). By a main void we mean the one where
electrons providing the conductivity of the metal are found. Under the
conditions of extremely strong anomalous skin effect such metals in the single
crystal state, as well as in the polycrystalline state, must have the impedance
depending on the electron mean free path (see Eq.(45)). In this case we also
must pay attention to the unusual relation between the imaginary and real parts
of the effective impedance. Possibly just this fact will be helpful for the
search of metals with the impedance (45).   
 
One of the most important phenomena related to the geometry of the Fermi
surface is the electronic topological transition manifesting itself in the
change of electronic properties of the metal due to the change of the topology
of the Fermi surface \cite{21}. In fact, all properties of the metal are
"sensitive" to the topological transition (see ref. \cite{22}), but when
characteristics of the phenomenon do not depend on the electron mean free path,
the influence of the topological transition is particularly clear. Extremely
anomalous skin effect is one of such phenomena. Our analysis shows that
polycrystals also have singularities of the effective impedance due to the
topological transition (the singularities "survive" in polycrystals). When the
value of the impedance of the single crystal metal is not very sensitive to the
orientation of the crystallographic axes with respect to the metal surface, the
character of the singularity of the effective impedance of the polycrystal is
the same as in the single crystal. If the value of the impedance of the single
crystal metal depends on the orientation of the crystallographic axes
essentially (as it takes place for the formation of a neck of the Fermi
surface, for example, of the goffered cylinder type), the character of the
singularity can change (see Eqs.(53) and Eqs.(54)).

The results of Sections IV and V confirm the following conclusion: if the
anisotropy of the Fermi surface is essential, the averaging necessary when
calculating the effective impedance of polycrystals does not liquidate the
influence of the geometry of the Fermi surface. In other words, it is not
sufficient to think about a polycrystal as of a metal with an effective
spherical Fermi surface, since in this case, some characteristic features of
extremely anomalous skin effect in polycrystals can be missed.  
\vskip 2mm
\centerline{\bf ACKNOWLEDGEMENT}
\vskip 2mm

The authors are grateful to Professor A.M.Dykhne in cooperation with whom the
results of Section II have been obtained. We would also like to thank him for
helpful comments during the course of this work. Also we are grateful to
M.Litinskaia for her help in preparation of this manuscript to publication. The
work of I.M.K. was supported by RBRF Grant No. 99-02-16533.
\vskip 2mm
\centerline{\bf APPENDIX 1}
\vskip 2mm

Let $\gamma$ denote a rotation about the three Euler angles 
${\theta}_{k},{\varphi}_{k},{\psi}_{k}$ which transforms the set of laboratory 
unit vectors ${\bf e}_{i}$ into the set of crystallographic unit vectors 
${\bf a}_{i}$. There are some different ways of the Euler angles definition
\cite{31}. To define concretely the elements of the rotation matrix (18), we
have to specify the definition of the Euler angles used in the present paper.
We assume that the set of crystallographic unit vectors is obtained from 
the set ${\bf e}_{i}$ in the following way: 

1.We rotate the vector ${\bf e}_{1}$ about the angle ${\varphi}_{k}$ at the 
plane $(1,2)$. The rotation results in the set of unit vectors ${\bf e'}_{i}$. 

2.Then we rotate the vector ${\bf e'}_{3}$ about the angle ${\theta}_{k}$ at 
the plane $(1',3')$ and obtain the set of unit vectors ${\bf e''}_{i}$. 

3.The last rotation which transfers the vectors ${\bf e''}_{i}$ into
the vectors ${\bf a}_{i}$ is the rotation of the vector ${\bf e''}_{i}$ 
about the angle ${\psi}_{k}$ at the plane $(1'',2'')$. 

Then the elements of the rotation matrix are:
$$
{\alpha}_{11} = [\cos{\theta}_{k}\cos{\varphi}_{k}\cos{\psi}_{k} -
\sin{\varphi}_{k}\sin{\psi}_{k}];
$$
$$
{\alpha}_{12} = -[\cos{\theta}_{k}\cos{\varphi}_{k}\sin{\psi}_{k} +
\sin{\varphi}_{k}\cos{\psi}_{k}];
$$
$$
{\alpha}_{21} = [\cos{\theta}_{k}\sin{\varphi}_{k}\cos{\psi}_{k} +
\cos{\varphi}_{k}\sin{\psi}_{k}];
$$
$$
{\alpha}_{22} = [-\cos{\theta}_{k}\sin{\varphi}_{k}\sin{\psi}_{k} +
\cos{\varphi}_{k}\cos{\psi}_{k}];
$$
$$
{\alpha}_{13} = \sin{\theta}_{k}\cos{\varphi}_{k};\qquad
{\alpha}_{31} = -\sin{\theta}_{k}\cos{\psi}_{k}; 
$$
$$
{\alpha}_{23} = \sin{\theta}_{k}\sin{\varphi}_{k};\qquad
{\alpha}_{32} = \sin{\theta}_{k}\sin{\psi}_{k};
$$
$$
{\alpha}_{33} = \cos{\theta}_{k}.    \eqno (A1.1)
$$

The averaging over the three Euler angles is defined as follows
$$
<...> = \frac{1}{8{\pi}^{2}}\int_0^{\pi} \sin {\theta}_{k}d{\theta}_{k}
\int_0^{2\pi} d{\varphi}_{k}\int_0^{2\pi} ... d{\psi}_{k}.  \eqno (A1.2)
$$
\vskip 2mm
\centerline{\bf APPENDIX 2}
\vskip 2mm

Let the Fermi surface of a single crystal metal be a surface of revolution, and
let the equation of this surface written with respect to the crystallographic
coordinate system be ${\varepsilon}_{F} = \varepsilon({p}_{\bot},{p}_{3})$. To
calculate the functions ${S}_{1,2}(\gamma)$, which enter Eq.(23.1) for the
local surface impedance, first of all we have to calculate the tensor
${S}_{\alpha\beta}(\gamma)$ defined by Eq.(17.2). Usually (see, for example,
ref. \cite{16}) the elements of this tensor are expressed as the functionals of
the Gaussian curvature of the points at the Fermi surface belonging to "the
belt" that is the geometric locus of the points where ${\bf kv}=0$. For our
purpose it is more convenient to write this tensor in a different form.

We start with writing the elements of the tensor $\hat S$ with respect to
crystallographic axes. We use the polar coordinates ${p}_{\bot}, \phi$ to
calculate the integrals in the expression (17.2) for ${S}_{ik}^{(a)}$. Since
with respect to the crystallographic axes the electron velocity ${\bf v}^{(a)}
= ({v}_{\bot}\cos \phi, {v}_{\bot}\sin \phi,{v}_{z})$ and the wave vector 
${\bf k}^{(a)} = k(-\sin {\theta}_{k}\cos {\psi}_{k},
\sin {\theta}_{k}\sin {\psi}_{k},\cos {\theta}_{k})$, we have 
$\delta ({\bf kv}/kv) = (v/{v}_{\bot}\sin {\theta}_{k})
\delta ({v}_{z}\cot {\theta}_{k}/{v}_{\bot} - \cos ({\psi}_{k}+\phi))$. 
Now the expressions for the elements of the tensor ${S}_{ik}^{(a)}$ are
$$
{\sigma}_{ik}^{(a)}({\theta}_{k},{\psi}_{k}) =
\frac{4{\sigma}_{a}}{{S}_{F}\sin {\theta}_{k}} \int \frac{{v}_{i}{v}_{k}}
{{v}_{\bot}|{v}_{z}|}\delta (\frac{{v}_{z}}{{v}_{\bot}}\cot {\theta}_{k} -
\cos (\phi + {\psi}_{k}))\quad {p}_{\bot}d{p}_{\bot}d\phi.    \eqno (A2.1)
$$
Carrying out the integration with respect to the angle $\phi$, we obtain the
elements of the tensor ${\hat S}^{(a)}(\gamma)$: 
$$
{S}_{11}^{(a)}(\gamma ) = \frac{8}{{S}_{F}\sin{\theta}_{k}}\{ {F}_{1} + 
\cos 2{\psi}_{k}[2\cot^2{\theta}_{k}{F}_{2} - {F}_{1}]\};    \eqno (A2.2a)
$$
$$
{S}_{12}^{(a)}(\gamma ) = - \frac{8\sin 2{\psi}_{k}}{{S}_{F}\sin {\theta}_{k}}
[2\cot^2{\theta}_{k}{F}_{2} - {F}_{1}];      \eqno (A2.2b)
$$
$$
{S}_{13}^{(a)}(\gamma ) = \frac{16\cos{\psi}_{k}\cot{\theta}_{k}}
{{S}_{F}\sin{\theta}_{k}}{F}_{2};    \eqno (A2.2c)
$$
$$
{S}_{22}^{(a)}(\gamma ) = \frac{8}{{S}_{F}\sin{\theta}_{k}}\{ {F}_{1} - 
\cos 2{\psi}_{k}[2\cot^2{\theta}_{k}{F}_{2} - {F}_{1}]\};    \eqno (A2.2d)
$$
$$
{S}_{23}^{(a)}(\gamma ) = -\frac{16\sin{\psi}_{k}\cot{\theta}_{k}}
{{S}_{F}\sin{\theta}_{k}}{F}_{2};    \eqno (A2.2e)
$$
$$
{S}_{33}^{(a)}(\gamma ) = \frac{16}{{S}_{F}\sin{\theta}_{k}}{F}_{2};  
\eqno (A2.2f) 
$$
Here the functions ${F}_{1}({\theta}_{k})$ and ${F}_{2}({\theta}_{k})$ are:
$$
{F}_{1}({\theta}_{k}) = \int\frac{{v}^{2}_{\bot}{p}_{\bot}d{p}_{\bot}}
{{v}_{z}\sqrt{{v}_{\bot}^{2}-v_z^2 \cot^2 {}\theta_{k}}},   \eqno (A2.3a)
$$
$$
{F}_{2}({\theta}_{k}) = \int\frac{{v}_{z}{p}_{\bot}d{p}_{\bot}}
{\sqrt{{v}_{\bot}^{2}-v_z^2 \cot^2 {}\theta_{k}}}.   \eqno (A2.3b)
$$
The integration is carried out over the region of the Fermi surface inside one
cell of the momentum space, where  $v_z > 0$ and the radicand under the
integral sign is positive.  

When Eqs.(A1.1) and (A2.2) are used to calculate the elements of the tensor 
${S}_{ik}(\gamma )={\alpha}_{ip}(\gamma ){\alpha}_{kq}(\gamma )
{S}_{pq}^{(a)}(\gamma )$ with respect to the laboratory coordinate system we
obtain: 
$$
{S}_{11}(\gamma ) = \frac{8}{{S}_{F}\sin{\theta}_{k}}[F({\theta}_{k}) - 
\cos 2{\varphi}_{k}\Phi ({\theta}_{k})];     \eqno (A2.4a)
$$
$$
{S}_{12}(\gamma ) = -\frac{8\sin 2{\varphi}_{k}}{{S}_{F}\sin{\theta}_{k}} 
\Phi ({\theta}_{k});     \eqno (A2.4b)
$$
$$
{S}_{22} = \frac{8}{{S}_{F}\sin{\theta}_{k}}[F({\theta}_{k}) + 
\cos 2{\varphi}_{k}\Phi ({\theta}_{k})];     \eqno (A2.4c)
$$
and
$$
{S}_{i3}(\gamma ) = 0; \qquad i=1,2,3.       \eqno (A2.4d)
$$
Here we introduced the functions
$$
F({\theta}_{k}) = {F}_{1}({\theta}_{k}) + {F}_{1}({\theta}_{k}); 
$$
$$
\Phi ({\theta}_{k}) = {F}_{1}({\theta}_{k}) - \frac{(1+\cos^2 {\theta}_{k})}
{\sin^2 {\theta}_{k}}{F}_{2}({\theta}_{k}).  \eqno (A2.5)
$$

We use Eqs.(A2.4) to calculate the functions ${S}_{\pm}(\gamma )$, $s(\gamma )$
and $R(\gamma )$ which enter Eqs.(22) for the local surface impedance. We
obtain: 
$$
{S}_{+}(\gamma ) = \frac{16}{{S}_{F}\sin {\theta}_{k}}F({\theta}_{k}), \qquad
{S}_{-}(\gamma ) = -\frac{16\cos 2{\varphi}_{k}}
{{S}_{F}\sin {\theta}_{k}}\Phi ({\theta}_{k}),   \eqno (A2.6a)
$$
$$
R(\gamma ) = \frac{16}{{S}_{F}\sin {\theta}{k}}\Phi ({\theta}_{k}), \qquad
s(\gamma ) = - \cos 2{\varphi}_{k}.   \eqno (A2.6b)
$$

The expressions for the functions ${S}_{1,2}({\theta}_{k})$ (see Eqs.(25)) 
follow directly from the substitution of Eqs.(A2.6) into Eqs.(23).
\vskip 2mm
\centerline{\bf APPENDIX 3}
\vskip 2mm 

Here we present the formulae, useful when calculating the effective impedance
of polycrystals composed of the grains whose Fermi surface is of the goffered
cylinder type (47). We start with the definition of the domain of integration
in Eqs.(50.4). When Eqs.(50.4) are used to calculate ${\tilde S}_{1(2)}$
we must have in mind that the interval of integration depends both on $w$ and
$\delta\varepsilon$. If we introduce 
$$
{x}^{2}_{\pm} = \frac{2}{\gamma}[(1-w)+\delta\varepsilon\pm 
\Delta (w;\delta\varepsilon)];\quad
\Delta (w;\delta\varepsilon) = \sqrt{{(1-w)}^{2} - 2\delta\varepsilon w},
\eqno (A3.1)
$$
it can be shown that:

\noindent 
1. if $\delta\varepsilon < 0$ for all $w$ ,
$$
{x}_{+} < x < \sqrt{\frac{2}{\gamma}(2+\delta\varepsilon)};  \eqno (A3.2a)
$$

\noindent 
2. if $\delta\varepsilon = 0$, 
$$
2\sqrt{\frac{1-w}{\gamma}} < x < \frac{2}{\sqrt{\gamma}},\;{\rm when}\;w < 1,
\qquad 0 < x < \frac{2}{\sqrt{\gamma}},\;{\rm when}\;w > 1;   \eqno (A3.2b)
$$

\noindent 
3. if $\delta\varepsilon > 0$,
$$
0< x < {x}_{-} \quad {\rm and}\quad {x}_{+} < x < 
\sqrt{\frac{2}{\gamma}(2+\delta\varepsilon)},\quad {\rm when}\; 
0 < w < {w}_{-};
$$
$$
\sqrt{\frac{2}{\gamma}\delta\varepsilon} < x <
\sqrt{\frac{2}{\gamma}(2+\delta\varepsilon)},\quad 
{\rm when}\quad {w}_{-}<w<{w}_{+};
$$
$$
{x}_{+} < x < \sqrt{\frac{2}{\gamma}(2+\delta\varepsilon)},\quad{\rm when}\quad
{w}_{+} < w.       \eqno (A3.2c)
$$ 
Here ${w}_{\pm} = 1+\delta\varepsilon \pm \sqrt{\delta\varepsilon(2 + 
\delta\varepsilon)}$.

\underline{Let's begin with the case $\delta\varepsilon < 0$}. When calculating
the functions $\Delta {\tilde S}_{\alpha}(w;\delta\varepsilon)$ (see Eq.(50.4)
and Eq.(52.3)), the cases $w<1$ and $w>1$ have to be examined separately. The
point is that when $\delta\varepsilon \ll 1$, the expansions of the function
${\Delta}^{(<)}(w;\delta\varepsilon)$ in the vicinity of the point
$\delta\varepsilon = 0$ are different for $w<1$ and $w>1$. When $0<w<1$, we
write the expression for ${\Delta}^{(<)}(w<1;\delta\varepsilon)$ as 
$$
{\Delta}^{(<)}(w<1;\delta\varepsilon) = \sqrt{{(1-w)}^{2} + 
2w|\delta\varepsilon |} = (1-w) + 2\delta (w;|\delta\varepsilon |), 
\eqno (A3.3a)
$$
where $\delta (w;|\delta\varepsilon |)\le\sqrt{2|\delta\varepsilon |}$ for all
$w<1$ and $|\delta\varepsilon | \ll 1$. We showed that up to the terms of the 
order of $\delta$  
$$
\Delta {\tilde S}_{1}^{(<)} (w<1) = \pi\frac{\sqrt{1+\gamma w}}
{{\gamma}^2 w} \delta (w;|\delta\varepsilon |)\sqrt{1-w},  \eqno (A3.3b)
$$
$$
\Delta {\tilde S}_{2}^{(<)} (w<1) = \pi 
\frac{{(1+\gamma w)}^{3/2}}{{\gamma}^2 w}\frac{\delta (w;|\delta\varepsilon |)}
{\sqrt{1-w}}, \eqno (A3.3c)
$$

When $w>1$, we set
$$
{\Delta}^{(<)}(w>1) = \sqrt{{(w-1)}^{2}+2w|\delta\varepsilon |}
= (w-1) + 2{\eta}^{(<)}(w;|\delta\varepsilon |), \eqno (A3.4a)
$$
where ${\eta}^{(<)}(w;|\delta\varepsilon |) \ll 1$ for all $w>1$ and 
$|\delta\varepsilon | \ll 1$: $|\delta\varepsilon |/2 \le 
{\eta}^{(<)}(w;|\delta\varepsilon |) \le \sqrt{|\delta\varepsilon |/2}$. For 
$\Delta {\tilde S}_{1(2)}^{(<)} (w>1)$ up to the leading terms in 
${\eta}^{(<)}$ we have  
$$
\Delta {\tilde S}_{1}^{(<)} (w>1) = -\frac{1}{{\gamma}^2}
\sqrt{\frac{1+\gamma w}{w}}\sqrt{\frac{w-1}{w}}
{\eta}^{(<)}(w;|\delta\varepsilon |)\ln {\eta}^{(<)}
(w;|\delta\varepsilon |), \eqno (A3.4b)
$$
$$
\Delta {\tilde S}_{2}^{(<)} (w>1) = \frac{1}{{\gamma}^2}
{\left( \frac{1+\gamma w}{w}\right)}^{3/2}\sqrt{\frac{w}{w-1}}
{\eta}^{(<)}(w;|\delta\varepsilon |)\ln {\eta}^{(<)} (w;|\delta\varepsilon |).
\eqno (A3.4c)
$$

We used Eqs.(A3.3) and Eqs.(A3.4) to obtain the functions 
$\delta{z}_{1(2)}^{(<)}(\delta\varepsilon)$ defined by Eq.(52.2). When 
calculating we divided the region of integration into two intervals: $0 < w <
1$ and $1 < w < \infty$. We showed that the leading term in 
$|\delta\varepsilon |$ of the sum $\delta{z}_{1}^{(<)}(\delta\varepsilon) + 
\delta{z}_{2}^{(<)}(\delta\varepsilon)$ which entered the expression (52.1)
for $\delta{\zeta}_{ef}^{(<)}(\delta\varepsilon)$, was defined by the interval 
$1 < w < \infty$. The result of the calculation is given by Eqs.(53).

\underline{Now we start with the case ${\varepsilon}_{F}/{\varepsilon}_{c}
> 1$ that is $\delta\varepsilon > 0$.} Here, according to Eq.(A3.2c), three
intervals of $w$ variation, namely: $0 < w < {w}_{-}$, ${w}_{-} < w < {w}_{+}$
and ${w}_{+} < w < \infty$, have to be examined separately. We showed that 
when $\delta\varepsilon \ll 1$, only $w$ from the interval ${w}_{+} < w < 
\infty$ contributed to the leading term of the expression for 
$\delta{\zeta}_{ef}^{(>)}(\delta\varepsilon)$.

For ${w}_{+} < w$ we introduced ${\eta}^{(>)}(w;\delta\varepsilon) \ll 1$
according to the formula  
$$
{\Delta}^{(>)} (w>{w}_{+};\delta\varepsilon) = \sqrt{{(w-1)}^{2}-
2w\delta\varepsilon } = (w-1) - 2{\eta}^{(>)}(w;\delta\varepsilon), 
\eqno (A3.5a)
$$
(compare with Eq.(A3.4a)). The leading terms in $\delta\varepsilon$ of the
expansions of $\Delta {\tilde S}_{1(2)}^{(>)} 
(w>{w}_{+})$ were given by the equations (A3.b) and (A3.c) where 
${\eta}^{(>)}(w;|\delta\varepsilon |)$ was substituted for 
${\eta}^{(<)}(w;\delta\varepsilon)$ and the signs changed. In this case the
calculation of the functions $\delta{z}_{1(2)}^{(>)} (\delta\varepsilon)$ was 
reduced to the calculation of the integral (52.2) over the the region
${w}_{+} < w < \infty$. The difference in the numerical factors 
${\alpha}^{(<)}$ and ${\alpha}^{(>)}$ entering the expressions (53.1) and
(54.1) for $\delta{\zeta}_{ef}^{(<)}$ and $\delta{\zeta}_{ef}^{(>)}$
respectively, arises from this last integration.
\vskip 2mm
\centerline{\bf APPENDIX 4}
\vskip 2mm

When Eqs.(25) - (27) are used to calculate the effective impedance of a
polycrystal composed of single crystal grains whose energy spectrum is defined
by Eq.(55) (the wurzite type single crystals), it is convenient to use 
$$
f({\theta}_{k}) = \frac{\cot {\theta}_{k}}
{\sqrt{{m}_{z}/{m}_{\bot}+\cot^2 {\theta}_{k}}}.  \eqno (A4.1a)
$$
in place of the Euler angle ${\theta}_{k}$. Then the averaging in Eq.(27.1)
corresponds to integration with respect to $f$: 
$$
<...> = \sqrt{\frac{m_z}{m_\bot}}\int\limits_0^1 \frac{(...)df}
{{[1+f^2(m_z/m_\bot - 1)]}^{3/2}}.  \eqno (A4.1b)
$$
Also in place of the momentum ${p}_{\bot}$ it is convenient to introduce the
dimensionless variable $x = ({p}_{\bot} -{p}_{0})/
\sqrt{2{m}_{\bot}{\varepsilon}_{F}}$. 

\underline{When $0< {\varepsilon}_{F} < {\varepsilon}_{c}^{+}$,} the Fermi
surface is the toroid ${\varepsilon}_{F} = {\varepsilon}^{-}({\bf p})$. If the
momentum ${p}_{\bot}$ is on the Fermi surface, $x$ belongs to the interval
$-1<x<1$. Next, the function $\Phi ({p}_{\bot}, \tan {\theta}_{k})$, which
enters the integrands of Eqs.(25), can be written as the function of $x$ and
$f$: 
$$
\Phi (x,f) = \frac{1}{f}\sqrt{\frac{x^2 - f^2}{1-x^2}}.  \eqno (A4.2)
$$
The radicand of Eq.(A4.2) is positive when $f^2 < x^2 <1$. Since $f<1$, the
last inequality defines the domain of integration in Eqs.(25). After these
comments our result for the effective impedance ${\zeta}_{ef}^{(<)}$,
Eq.(58.1), can be easily obtained. The function $B(z)$ in Eq.(58.1) is
$$
B(z) = \frac{{z}^{1/3}}{4}\int\limits_0^1 
\frac{df}{{[1+f^2(z-1)]}^{5/3}}\left\{B_1(f) +
{\left[\frac{z}{1+f^2(z-1)}\right]}^{1/3}B_2(f)\right\},   \eqno (A4.3a)
$$
and
$$
B_1(f) = {\left[\frac{E(\sqrt{1-f^2})-f^2K(\sqrt{1-f^2})}{1-f^2}\right]}
^{-1/3}; \quad B_2(f)={\left[\frac{K(\sqrt{1-f^2})-E(\sqrt{1-f^2})}
{1-f^2}\right]}^{-1/3};   \eqno (A4.3b)
$$
$K(k)$ and $E(k)$ are full elliptic integrals of the first and the second kind
respectively.

\underline{When ${\varepsilon}_{F} > {\varepsilon}_{c}^{+}$,} the Fermi surface
consists of the external toroidal part (${\varepsilon}_{F}=
{\varepsilon}^{-}({\bf p})$) and the internal ovaloid part (${\varepsilon}_{F}
= {\varepsilon}^{+}({\bf p})$). It worth to be mentioned, that in this case, in
Eqs.(25) not only an additional domain of integration related to the internal
part of the Fermi surface appears, but  the domain of integration related to
the external part of the Fermi surface also changes.

Calculating the functions ${S}_{1,2}$ according to Eqs.(25), we used
the defined above variable $x$ when integration was carried out over the
toroidal part of the Fermi surface. We introduced $y = ({p}_{\bot} +
{p}_{0})/\sqrt{2{m}_{\bot}{\varepsilon}_{F}}$, when integrating over its
ovaloid part. It can be easily seen that the domains of integration are:
$$
-\sqrt{{\varepsilon}_{c}^{+}/{\varepsilon}_{F}} <x < -f\; {\rm and}\; 
f < x < 1,\qquad {\rm if}\; 0 < f < \sqrt{{\varepsilon}_{c}^{+}
/{\varepsilon}_{F}}
$$
$$
f < x < 1,\qquad {\rm if}\; \sqrt{{\varepsilon}_{c}^{+}/{\varepsilon}_{F}} 
< f <1;  \eqno (A4.4a)
$$
and
$$
\sqrt{{\varepsilon}_{c}^{+}/{\varepsilon}_{F}} < y < 1,\qquad {\rm if} \; 
0 < f < \sqrt{{\varepsilon}_{c}^{+}/{\varepsilon}_{F}};
$$
$$
f < y < 1,\qquad {\rm if}\;\sqrt{{\varepsilon}_{c}^{+}/{\varepsilon}_{F}}< f<1.
\eqno (A4.4b)
$$

We used Eq.(58.1) and Eqs.(A4.3) to calculate ${\zeta}_{ef}^{(>)}
({\varepsilon}_{F}/{\varepsilon}_{c}^{+})$ for an arbitrary value of the
parameter $\delta\varepsilon =({\varepsilon}_{F} - {\varepsilon}_{c}^{+})
/{\varepsilon}_{c}^{+}>0$. Near the point of the topological transition, $0 <
\delta\varepsilon \ll 1$, our result for $\Delta\zeta (\delta\varepsilon)$ is
given by Eq.(59.1), where the function $C(z)$ is
$$
C(z) = \frac{{z}^{1/3}}{36}\int\limits_0^1
\frac{{[{B}_{1}(f)]}^{4} df}{{[1+f^2(z-1)]}^{5/3}\sqrt{1-f^2}}. \eqno (A4.5)
$$
The function $B_1(f)$ is again given by Eq.(A4.3b).

\newpage
\centerline{\bf List of Figures:}
\vskip 2mm
\noindent
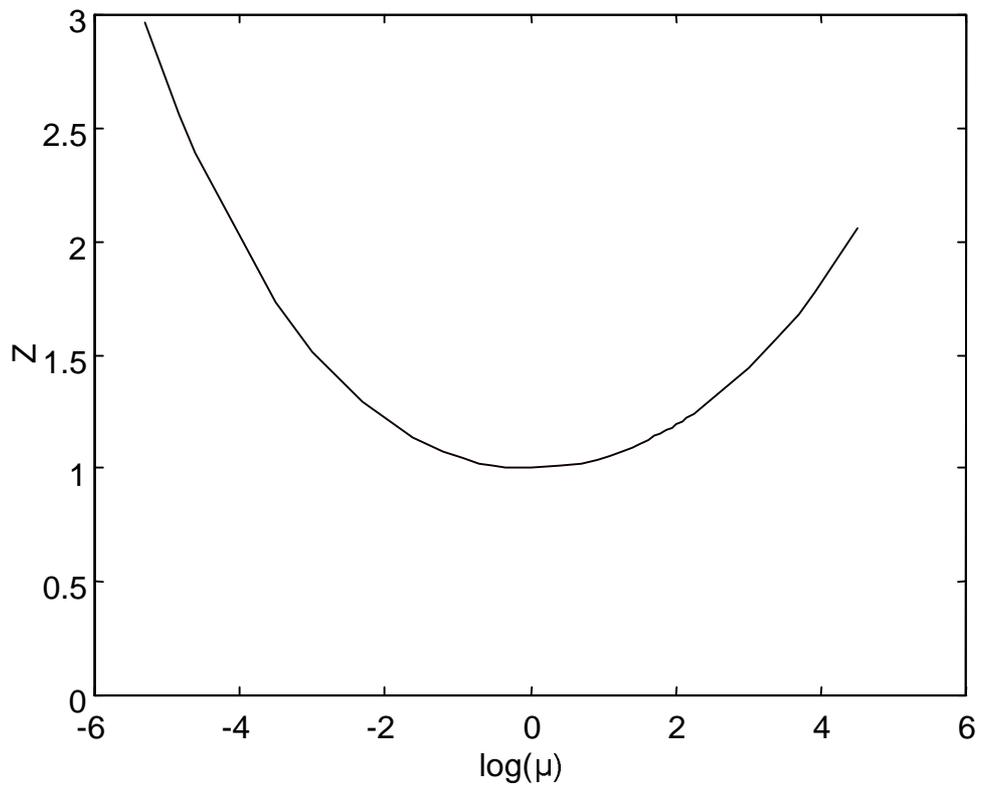
Fig.1. The function $Z = Z(\mu)$ defined by Eq.(32.2).
\vskip 2mm
\noindent
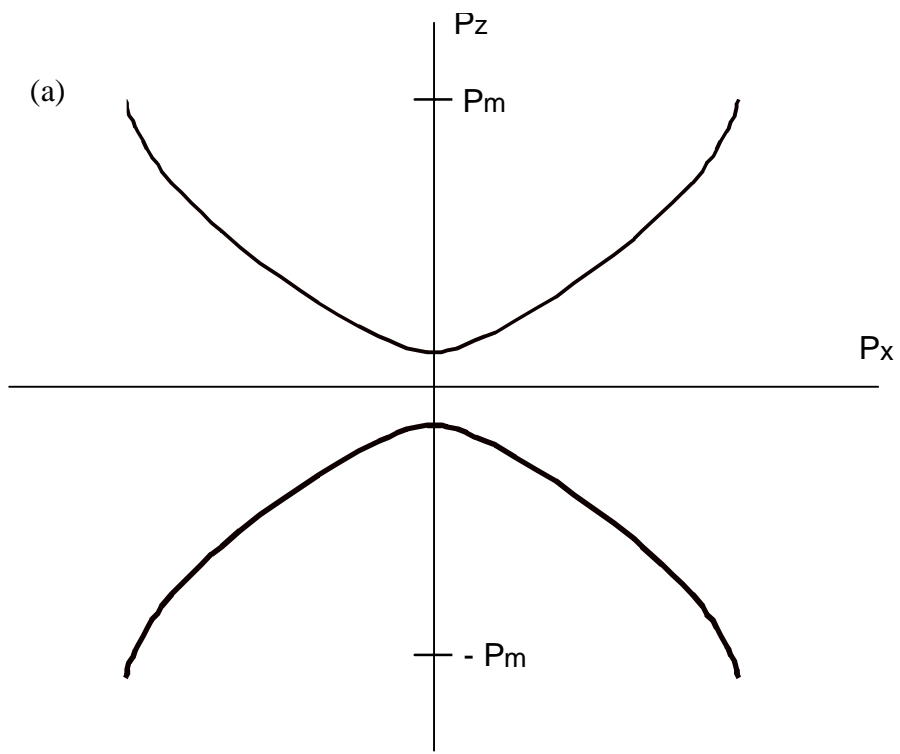
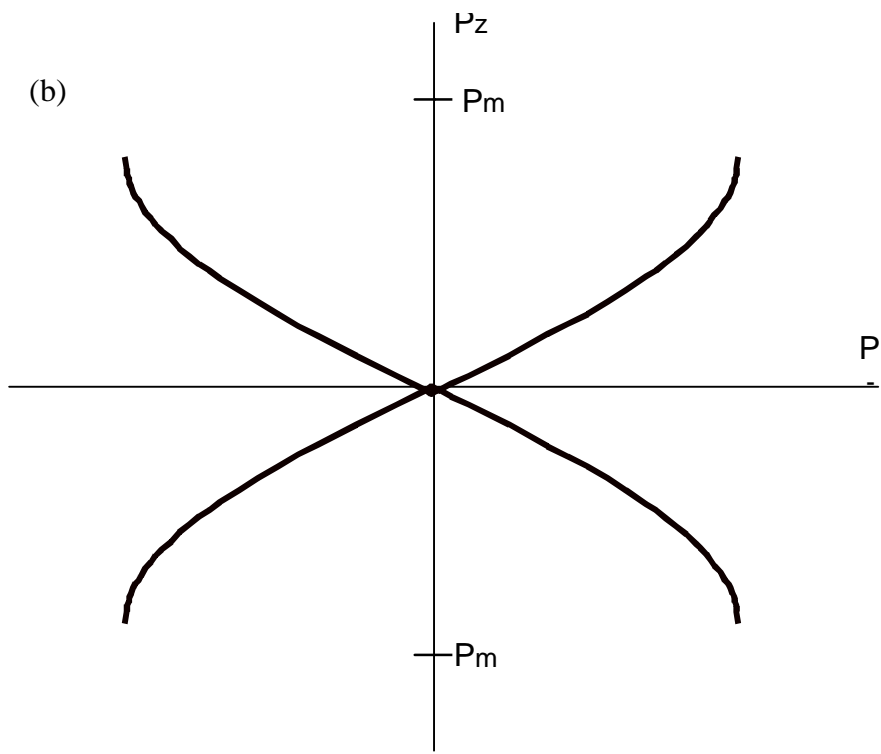
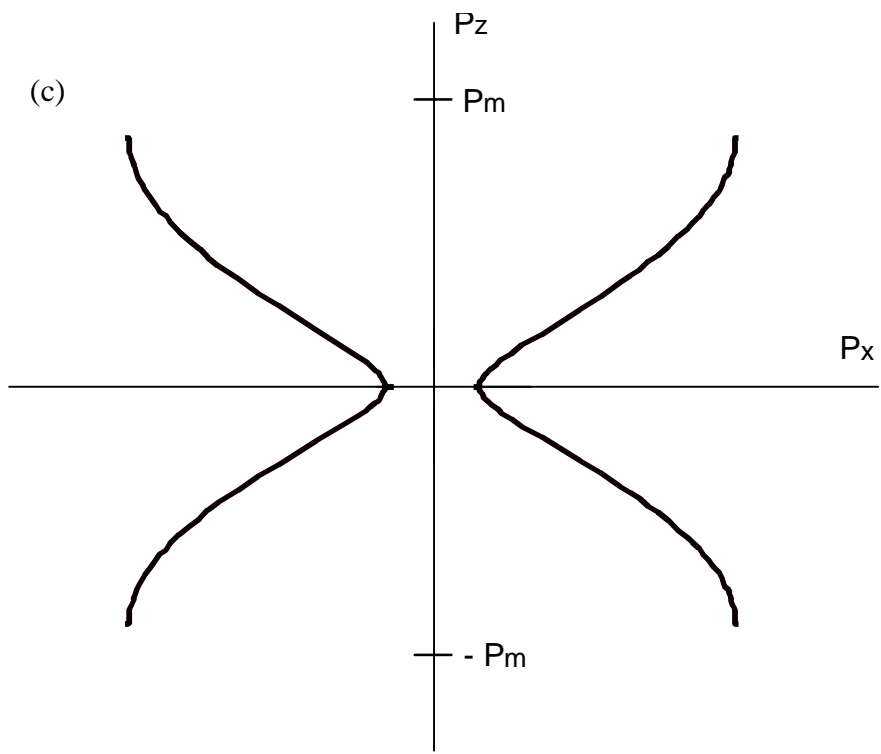
Fig.2. The crossection of the Fermi surface of a goffered cylinder type:
a) ${\varepsilon}_{F} < {\varepsilon}_{c}$, the Fermi surface is a closed
surface; b) ${\varepsilon}_{F} = {\varepsilon}_{c}$, the Fermi surface has a
conic point; c) ${\varepsilon}_{F} > {\varepsilon}_{c}$, the Fermi surface is
an open surface.
\vskip 2mm
\noindent
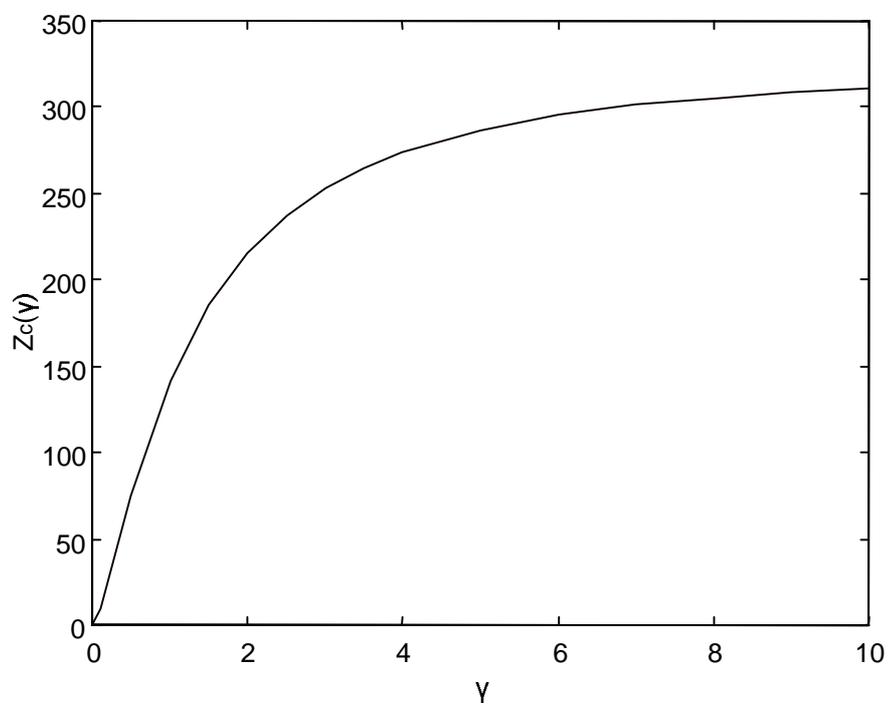
Fig.3. The function $Z = {Z}_{c}(\gamma)$ defining the effective impedance when
the Fermi surface (47) has the conic point.
\vskip 2mm
\noindent
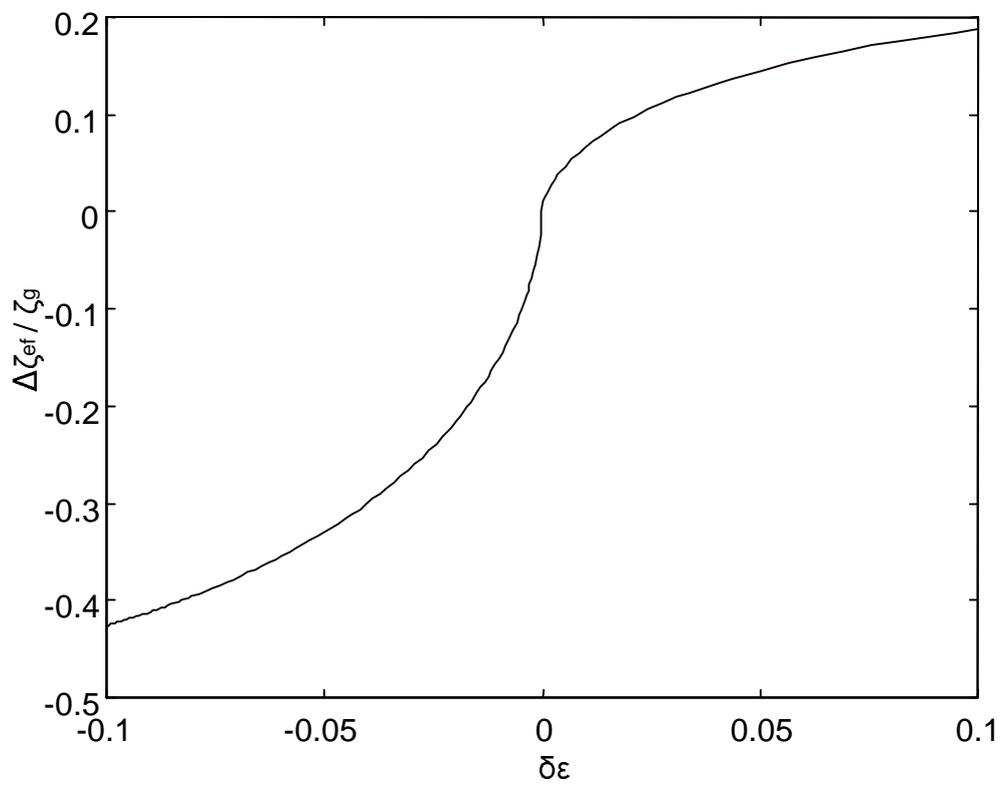
Fig.4. The singularity of the effective impedance in the vicinity of the
electronic topological transition for the case of the neck formation.
\vskip 2mm
\noindent
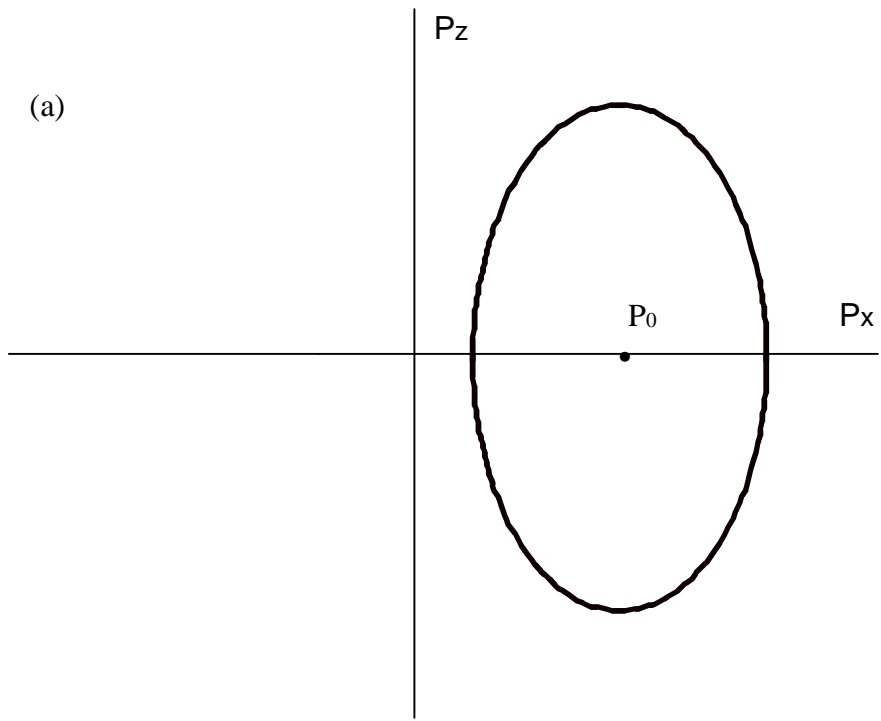
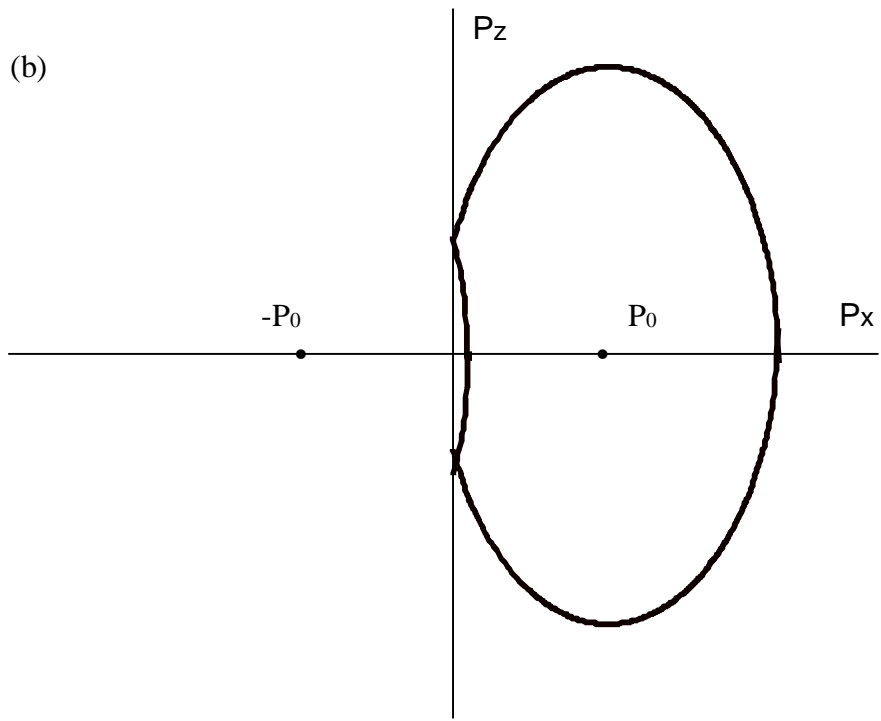
Fig.5. The equienergy surfaces (55) are obtained by rotation about the axis
${p}_{z}$ of the curves shown in the figure (a) for the energies $0<
\varepsilon < {\varepsilon}_{c}^{+}$ and (b) for the energies $\varepsilon > 
{\varepsilon}_{c}^{+}$.

\end{document}